\documentclass{aa}

\usepackage{graphicx}
\usepackage{txfonts}
\usepackage{amsmath}
\usepackage{threeparttable}

\usepackage[
bookmarks=true,           
bookmarksnumbered=true,   
colorlinks=true,          
citecolor=blue,           
linkcolor=blue,           
menucolor=blue,           
urlcolor=blue,            
linkbordercolor={0 0 1},  
pdfborder={0 0 1},
frenchlinks=true]{hyperref}

\providecommand{\adsurl}[1]{\href{#1}{ADS}}

\DeclareSymbolFont{UPM}{U}{eur}{m}{n}
\DeclareMathSymbol{\umu}{0}{UPM}{"16}
\let\oldumu=\umu
\renewcommand\umu{\ifmmode\oldumu\else\math{\oldumu}\fi}
\newcommand\micro{\umu}
\newcommand\micron{\micro\rm m}
\newcommand\microns{\micron}

\newcommand\HST{{\em HST}}
\newcommand\Hubble{{\em Hubble}}
\newcommand\Spitzer{{\em Spitzer}}

\newcommand\oneeightnine{HD~189733}
\newcommand\oneeightnineb{HD~189733\,b}

\newcommand\twoohnineb{HD~209458\,b}

\newcommand\rjup{\ifmmode{R\sb{\rm Jup}}\else$R$\sb{Jup}\fi}
\newcommand\mjup{\ifmmode{M\sb{\rm Jup}}\else$M$\sb{Jup}\fi}
\newcommand\kms{km\,s$^{-1}$}

\newcommand\msun{\ifmmode{M\sb{\odot}}\else$M\sb{\odot}$\fi}
\newcommand\rsun{\ifmmode{R\sb{\odot}}\else$R\sb{\odot}$\fi}
\newcommand\mearth{\ifmmode{M\sb{\oplus}}\else$M\sb{\oplus}$\fi}
\newcommand\rearth{\ifmmode{R\sb{\oplus}}\else$R\sb{\oplus}$\fi}
\newcommand\Rp{\ifmmode{R\sb{\rm p}}\else$R\sb{\rm p}$\fi}
\newcommand\rstar{\ifmmode{R\sb{\rm s}}\else$R\sb{\rm s}$\fi}
\newcommand\fstar{\ifmmode{F\sb{\rm s}}\else$F\sb{\rm s}$\fi}
\newcommand\rprs{\Rp/\rstar}
\newcommand\fspot{\ifmmode{f\sb{\rm spot}}\else$f\sb{\rm spot}$\fi}
\newcommand\Fspot{\ifmmode{F\sb{\rm spot}}\else$F\sb{\rm spot}$\fi}

\newcommand\tspot{\ifmmode{T\sb{\rm spot}}\else$T\sb{\rm spot}$\fi}
\newcommand\tstar{\ifmmode{T\sb{\rm eff}}\else$T\sb{\rm eff}$\fi}

\newcommand\mgi{\ion{Mg}{i}}
\newcommand\mgii{\ion{Mg}{ii}}
\newcommand\fei{\ion{Fe}{i}}
\newcommand\feii{\ion{Fe}{ii}}
\newcommand\hi{\ion{H}{i}}

\newcommand\molhyd{\ifmmode{{\rm H}\sb{2}}\else{H$\sb{2}$}\fi}
\newcommand\methane{\ifmmode{{\rm CH}\sb{4}}\else{CH$\sb{4}$}\fi}
\newcommand\water{\ifmmode{{\rm H}\sb{2}{\rm O}}\else{H$\sb{2}$O}\fi}
\newcommand\carbdiox{\ifmmode{{\rm CO}\sb{2}}\else{CO$\sb{2}$}\fi}
\newcommand\carbmono{\ifmmode{{\rm CO}}\else{CO}\fi}
\newcommand\ammonia{\ifmmode{{\rm NH}\sb{3}}\else{NH$\sb{3}$}\fi}
\newcommand\acetylene{\ifmmode{{\rm C}\sb{2}{\rm H}\sb{2}}
                        \else{C$\sb{2}$H$\sb{2}$}\fi}
\newcommand\ethylene{\ifmmode{{\rm C}\sb{2}{\rm H}\sb{4}} 
                        \else{C$\sb{2}$H$\sb{4}$}\fi}
\newcommand\cyanide{\ifmmode{{\rm HCN}}\else{HCN}\fi}
\newcommand\nitrogen{\ifmmode{{\rm N}\sb{2}}\else{N$\sb{2}$}\fi}

\newcommand\mcc{\textsc{mc3}}
\newcommand\pyratbay{\textsc{Pyrat Bay}}
\newcommand\repack{\textsc{repack}}

\begin{document} 

\title{The {\Hubble}/STIS Near-ultraviolet
  Transmission Spectrum of HD~189733\,b}
\titlerunning{The {\HST} NUV Transmission Spectrum of \oneeightnineb}

\author{
Patricio~E.~Cubillos\inst{1,2}
\and Luca~Fossati\inst{1}
\and Tommi~Koskinen\inst{3}
\and Chenliang~Huang\inst{3}
\and A.~G.~Sreejith\inst{1,4}
\and Kevin~France\inst{4}
\and P. Wilson Cauley\inst{4}
\and Carole~A.~Haswell\inst{5}
}
\institute{
Space Research Institute, Austrian Academy of Sciences,
Schmiedlstrasse 6, A-8042, Graz, Austria \\
\email{patricio.cubillos@oeaw.ac.at}
\and
INAF -- Osservatorio Astrofisico di Torino,
Via Osservatorio 20, 10025 Pino Torinese, Italy
\and
Lunar and Planetary Laboratory, University of Arizona,
1629 E. University Blvd., Tucson, AZ 85721, USA
\and
Laboratory for Atmospheric and Space Physics,
University of Colorado, 600 UCB, Boulder, CO 80309, USA
\and
School of Physical Sciences, The Open University,
Walton Hall, Milton Keynes MK7 6AA, UK
}

 
\abstract
{The benchmark hot Jupiter {\oneeightnineb} has been a key target to
  lay out the foundations of comparative planetology for giant
  exoplanets.  As such, {\oneeightnineb} has been extensively studied
  across the electromagnetic spectrum.
  Here, we report the observation and analysis of three transit light
  curves of {\oneeightnineb} obtained with {\Hubble}/STIS in the near
  ultraviolet, the last remaining unexplored spectral window to be
  probed with present-day instrumentation for this planet.
  The NUV is a unique window for atmospheric mass-loss studies owing
  to the strong resonance lines and large photospheric flux.
  Overall, from a low-resolution analysis ($R=50$) we found that the
  planet's near-ultraviolet spectrum is well characterized by a
  relatively flat baseline, consistent with the optical-infrared
  transmission, plus two regions at $\sim$2350 and $\sim$2600 {\AA}
  that exhibit a broad and significant excess absorption above the
  continuum.
  From an analysis at a higher resolution ($R=4700$), we found that the
  transit depths at the core of the magnesium resonance lines are
  consistent with the surrounding continuum.  We discarded the
  presence of {\mgii} absorption in the upper atmosphere at a
  $\sim$2--4$\sigma$ confidence level, whereas we could place no
  significant constraint for {\mgi} absorption.  
  These broad absorption features coincide with the expected location
  of {\feii} bands; however, solar-abundance hydrodynamic
  models of the upper atmosphere are not able to reproduce the
  amplitude of these features with iron absorption.  Such scenario
  would require a combination of little to no iron condensation in the
  lower-atmosphere, super-solar metallicities, and a mechanism to
  enhance the absorption features (such as zonal wind broadening). The
  true nature of this feature remains to be confirmed.
}

\maketitle

\section{Introduction}

The hot Jupiter {\oneeightnineb} is one of the earliest transiting
exoplanets to be detected \citep{BouchyEtal2005aaHD189733discovery},
being one of the closest transiting planets to
Earth.  {\oneeightnineb}'s large mass of $1.13~\mjup$, large radius of
$1.13~\rjup$, and short-period 2.2-day orbit
\citep{StassunEtal2017ajGaiaRadiiMasses} provide some of the most
favorable conditions for atmospheric characterization.  Consequently,
{\oneeightnineb} has become one of the most extensively observed
exoplanets, now constituting a benchmark target for atmospheric
characterization via follow-up observations and theoretical modeling.

From space, the {\it Hubble Space Telescope} (\HST) has been
instrumental in probing the optical-to-infrared transmission spectra of
this planet.  Collective observations with the Advanced Camera for
Surveys (ACS), the Space Telescope Imaging Spectrograph (STIS), and
the Wide Field Camera~3 (WFC3) have obtained nearly continuous spectra
over the 0.3--1.7 ${\microns}$ range
\citep[][]{PontEtal2008mnrasHD189733bHaze,
  SingEtal2011mnrasHD189733b_STIS, HuitsonEtal2012mnrasHD189bHSTstis,
  McCulloughEtal2014apjHD189733bHST}.  Complementary infrared (IR)
transmission photometry from the {\it Spitzer} Space Telescope between
3.6 and 8.0 $\microns$ \citep{DesertEtal2009apjHD189733b,
  AgolEtal2010apjHD189733b,
  KnutsonEtal2012apj3.6um4.5umPhaseHD189733b}, and posterior
re-analyses of the cumulative data \citep{PontEtal2013mnrasHD189733b,
  SingEtal2016natHotJupiterTransmission} have revealed a steep Rayleigh
scattering slope dominating the optical and near-infrared spectrum.
These results are consistent with the presence of high-altitude
atmospheric hazes, ruling out a clear atmosphere.

Ground-based facilities have further allowed the detection of several
spectral features; in particular, high-resolution observations have
lead to the detection of narrow absorption features above the haze
layer from the H Balmer lines \citep{JensenEtalapj2012HalphaHD189733b,
  CauleyEtal2015apjHD189733bHiResHydrogenBow} and the Na\,D lines
\citep{RedfieldEtal2008apjHD189733b,
  WyttenbachEtal2015aaHD189733bSodiumHARPS}.
These observations constrained the temperature of the thermosphere,
revealed strong wind velocities consistent with an eastward equatorial
jet \citep{LoudenWheatley2015apjHD189733bHiresSodiumWinds},
and detected a transit early ingress, possibly caused by a hydrogen
bow-shock ahead of the planet
\citep{CauleyEtal2015apjHD189733bHiResHydrogenBow}.
We note however that these detections have been contested in
subsequent studies
\citep[see][]{BarnesEtal2016mnrasExcessAbsorptionHD189733b,
  GuilluyEtal2020aaGianoHD189733bHelium}, arguing that the observed
signal could be stellar in nature, rather than planetary.
Similarly, stellar effects are also suspected to affect
the steep UV slope observed by {\HST} for this planet, although the magnitude
of such contribution is not fully understood
\citep[see, e.g.,][]{McCulloughEtal2014apjHD189733bHST,
RackhamEtal2019ajStellarHeterogeneityII}.

High-resolution spectroscopy studies of {\oneeightnineb} at
near-infrared wavelengths via the cross-correlation method have
yielded the detection of {\water} absorption features
\citep{BirkbyEtal2013mnrasHD189733bHighRes,
  BrogiEtal2018aaHD189733bGianoH2O} and CO
\citep{RodlerEtal2013mnrasHD189733bHiresCO,
  BrogiEtal2016apjHD189733bHiresTransmission}. These observations
further constrain atmospheric composition and suggest small wind
speeds of $\sim$2 {\kms} (at pressures around the
$10^{-2} - 10^{-3}$ bar levels).

Infrared observations of the day-side emission during secondary
eclipse has revealed a lack of thermal inversion and a nearly solar
{\water} abundance in the lower planetary atmosphere
\citep[][]{CrouzetEtal2014apjHD189733bHSTwfc3EmissionH2O,
  TodorovEtal2014apjHD189733bSpitzerEmissionIRS}.  Furthermore,
infrared photometry along the entire orbital phase has measured the
brightness temperature variation, constraining the energy budget and
the global circulation regime
\citep{KnutsonEtal2007nat8umPhaseHD189733b,
  KnutsonEtal2009apj24umPhaseHD189733b,
  KnutsonEtal2012apj3.6um4.5umPhaseHD189733b}.  These observations
unveiled an efficient energy redistribution of the incident stellar
irradiation and an eastward offset of the peak emission, relative to
the sub-stellar point.  As with the conclusions from optical
observations, these measurements are consistent with circulation
models exhibiting an eastward advection by an equatorial
super-rotating jet with wind speeds of the order of $\sim${\kms} at pressures
around the mbar level \citep[e.g.,][]{ShowmanEtal2009apjRadGCM,
  ShowmanEtal2013apjHotJupiterCirculation}.

\subsection{Upper-atmosphere Characterization of {\oneeightnineb}}

Given the close planet-to-star distance, it is expected that
{\oneeightnineb} exhibits atmospheric escape.  In fact, the host star,
{\oneeightnine}, is an active K-dwarf star \citep[radius of
$0.75 R_\odot$, mass of $0.79 M_\odot$, effective temperature of
$5052$~K,][]{StassunEtal2017ajGaiaRadiiMasses}, showing clear evidence
of activity and spots during transit events
\citep[e.g.,][]{SingEtal2011mnrasHD189733b_STIS,
  PontEtal2013mnrasHD189733b}, which makes it a particularly
interesting study case for star--planet interactions
\citep[e.g.,][]{PoppenhaegerEtal2013HD189733bChandra,
  PillitteriEtal2015apjHD189733fuvVariability,
  CauleyEtal2018ajStarPlanetInteractionHD189733b}.

Ly$\alpha$ transit observation in the far-ultraviolet with {\HST}/STIS
detected an extended hydrogen envelope that expands hydrodynamically
and escapes to space, powered by absorption of the high-energy stellar
flux \citep[X-ray and extreme
ultraviolet;][]{LecavelierEtal2012aaHD189733bEvapVariation,
  BourrierEtal2013aaHD189733bLyAlpha}.  Strong temporal variations and
the high velocity of the escaping gas suggest the presence of
significant star--planet interactions (consistent with the active
nature of the host star), which requires an acceleration mechanism on
top of stellar radiation pressure, such as the stellar wind.  Although
later, \citet{GuoBenJaffel2016apjEUVinfluence} found that the
Ly$\alpha$ variability can be simply explained by invoking a thermal
population of {\hi} in the planet's upper atmosphere, without a need for
complex dynamics involving stellar winds or radiation pressure. In
this case, changes to the star's EUV output would explain the
different absorption signals.
Far-ultraviolet observations with the {\HST} Cosmic Origins
Spectrograph have also led to the detection of \ion{O}{i} and
\ion{C}{ii} in the planetary upper atmosphere
\citep{BenJaffelBallester2013aaHD189733bOxygenHST}, indicating that
heavy atoms are possibly entrained and dragged upwards in the hydrogen
hydrodynamic outflow of the atmosphere.

The H$\alpha$ absorption feature detection by
\citet{JensenEtalapj2012HalphaHD189733b} indicates that the upper
atmosphere of {\oneeightnineb} is neither in radiative equilibrium nor
thermodynamic equilibrium, due to the large ultraviolet flux received
by its active host star.  In addition,
\citet{SalzEtal2018aaHeHD189733b},
\citet{ZhangEtal2022ajVariableHeliumHD189733b},
and \citet{GuilluyEtal2020aaGianoHD189733bHelium} detected a strong
absorption signal from the \ion{He}{i} triplet at 10830 $\AA$ by means
of ground-based high-resolution transmission spectroscopy with
CARMENES and GIARPS, respectively.  The observed He absorption favors
a compact helium atmosphere with a substantial column density and no
evidence of He escape. This interpretation of the He observations
would seem to be consistent with the aforementioned explanation of the
Lyman alpha transit observations by
\citet{GuoBenJaffel2016apjEUVinfluence}.

The many observations of {\oneeightnineb} across the electromagnetic
spectrum have enabled an in-depth characterization of the composition
and physical state of both the lower and upper atmosphere of the
planet.  However, until recently, the near-ultraviolet (NUV) window, 
probing both the exosphere and thermosphere, escaped the 
spectral coverage probed by these observations.
{\citet{KingEtal2021mnrasNUVxmmHD189733b}} analyzed 20 NUV photometric
transits of {\oneeightnineb} obtained from the XMM-Newton optical
monitor.  They detected a transit-depth signal consistent with that in
the optical, ruling out extended broad-band absorption towards or
beyond the Roche lobe.  However, to detect the NUV absorption
signature from individual metallic lines, higher spectral-resolution
observations are required, like those from {\HST}/STIS, with a
resolving power of $R \approx 30,000$
\citep[see][]{SingEtal2019ajWASP121bTransmissionNUV,
CubillosEtal2020ajHD209458bNUV}.

For the slighly more strongly irradiated planet {\twoohnineb},
\citet{CubillosEtal2020ajHD209458bNUV} detected {\feii} absorption in
a 100 {\AA}-wide range around 2370 {\AA}, lying beyond the planetary
Roche lobe, while finding no evidence for absorption by {\mgii}, neutral
magnesium ({\mgi}), nor neutral iron
({\fei}).  These results suggest that hydrodynamic escape is strong
enough to carry heavy atoms beyond the planetary Roche lobe; however,
condensate formation sequesters primarily Mg (rather than Fe) in the
lower atmosphere of the planet, consistent with microphysical
cloud-formation models \citep{GaoEtal2020natasAerosolComposition}.
For the ultra-hot Jupiter WASP-121b,
\citet{SingEtal2019ajWASP121bTransmissionNUV} reported
the detection of strong narrow NUV absorption features, consistent
with the singly ionized magnesium ({\mgii}) doublet and several lines of 
ionized iron ({\feii}), emanating from altitudes that exceed the Roche lobe
radius of the planet.  These results indicate that {\mgii} and {\feii}
species are not trapped into condensates at depth, and can
hydrodynamically escape the atmosphere.

To complete the spectral survey of this benchmark planet and aid the
quantitative assessment of the metallicity, ionization state, and
outflow rate of the upper atmosphere of {\oneeightnineb}, here we
analyzed 15 {\HST} orbits obtained from three STIS near-ultraviolet
transit light curves.  In Section \ref{sec:observations} we describe
the observations in more detail.  In Section \ref{sec:analysis} we
describe our data-analysis procedure.  In Section \ref{sec:spectra} we
present the main results from the NUV data analysis. In Section
\ref{sec:context} we place the observations in context by contrasting
them with theoretical models of the planet. In Section
\ref{sec:speculation} we speculate on the origin of the features seen
in the NUV observations.  Finally, in Section \ref{sec:conclusions} we
summarize our conclusions.

\section{Observations}
\label{sec:observations}

We observed three transits of the planet {\oneeightnineb}, obtained
with {\HST}, using the Space Telescope Imaging Spectrograph (STIS,
program GO \#15338, PI L. Fossati).  The observations were carried out
on 2018-03-10, 2018-09-19, and 2019-08-27 (visits 1, 2, and 3,
respectively); each one using the NUV Multi-Anode Microchannel Array
(NUV-MAMA) detector in ACCUM mode, E230M grating, and with an aperture of
0{\farcs}2~$\times$~0{\farcs}2.

Each transit observation (one {\HST} visit) consists of five {\HST}
orbits.  For each visit, the fourth orbit occurs during transit,
whereas the first three and last {\HST} orbits occur out of transit.
The orbits are broken into exposures (frames) of $\sim$332 seconds
each.  The first two visits contain nine frames per orbit (except for
the first orbit, which contains one fewer frame to accommodate
acquisition observations).  The third visit contains one fewer frame
per orbit, possibly due to a longer pointing adjustment time after
{\HST} returned to operations from a gyroscope failure.

\begin{figure*}
\centering
\includegraphics[width=\linewidth]{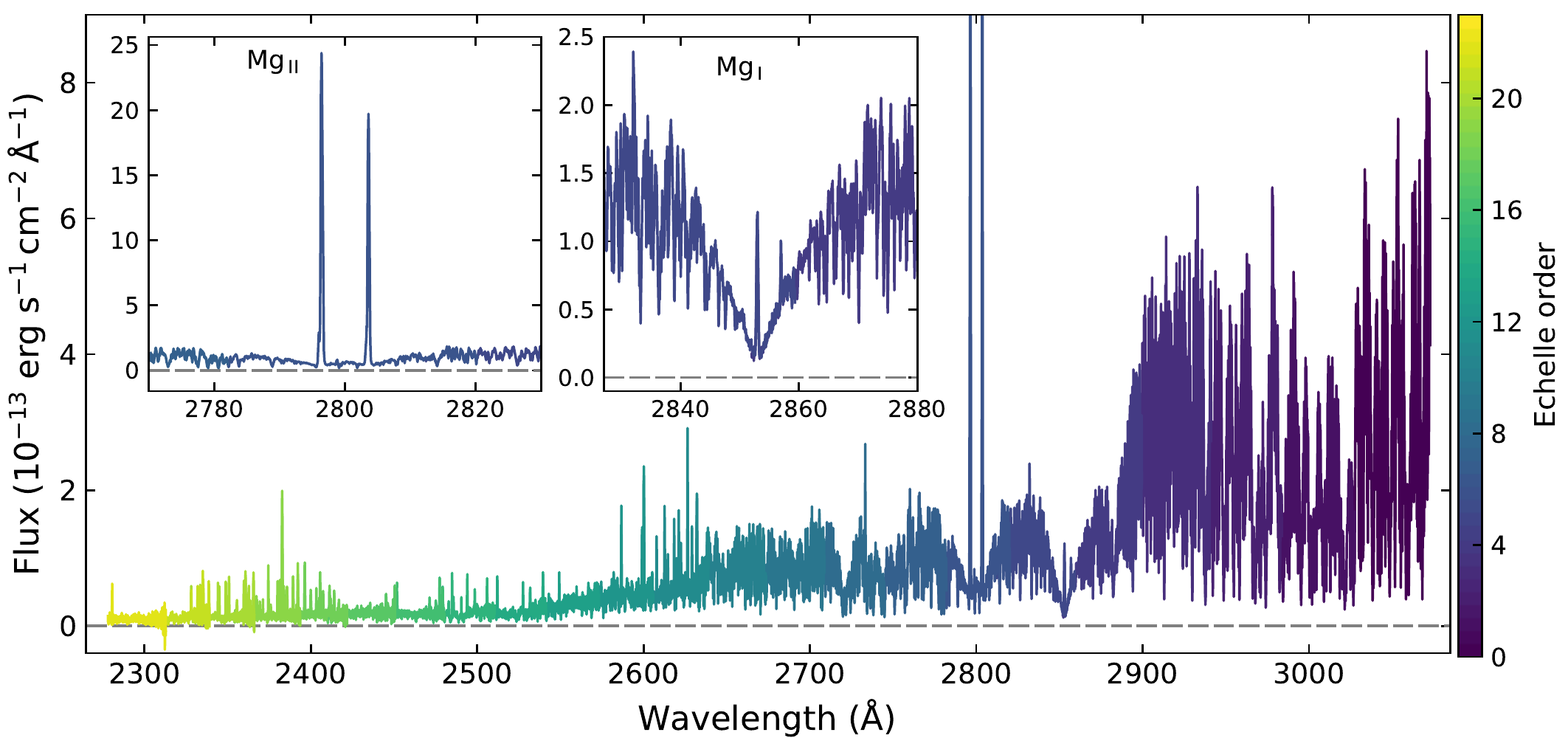}
\caption{
{\oneeightnine} system flux as observed by {\HST}/STIS.  This
  spectrum has been constructed from the mean time-series spectra obtained
  during the first visit, after discarding bad pixels.  The spectrum is
  color coded by echelle order (see color bar on the right).  The
  inset panels zoom-in around the location of the {\mgii} and {\mgi}
  resonance lines.}
\label{fig:stellar_spectrum}
\end{figure*}

Each frame consists of an echelle spectrum comprising 23 partially
overlapping orders, covering the 2300--3100 {\AA} spectral range.
Each order contains 1024 wavelength samples, with a resolving power of
$R = \lambda/\Delta\lambda = 30\,000$, where the resolution element
spans approximately two pixels (or equivalently $\sim$0.09 {\AA}, or
$\sim$10 {\kms}).  Figure \ref{fig:stellar_spectrum} shows a sample
spectrum from visit 1.  As expected, the flux decreases toward shorter
wavelengths, with a clear drop in signal-to-noise ratio (S/N) at wavelengths
shorter than $\sim2550\,\AA$.  The stellar spectrum shows many
emission features, most of them correspond to {\feii} lines rising from 
low energy levels, but also the {\mgi} and {\mgii} resonance lines 
(the strongest line-emission feature in the data,
as shown in the inset panels of Fig.\ \ref{fig:stellar_spectrum}).
In contrast, stellar {\fei} features appear in absorption.

The input for our analysis consisted of the x1d CALSTIS-reduced fits
files\footnote{\href{https://hst-docs.stsci.edu/stisdhb/chapter-3-stis-calibration}{https://hst-docs.stsci.edu/stisdhb/chapter-3-stis-calibration}}
(version 3.4.2).  The reduction comprises dark and flat-field
correction, 1D spectral extraction of each echelle order, background
subtraction, and wavelength and flux calibration.  The results are the
1D spectra (erg\,s$^{-1}$cm$^{-2}${\AA}$^{-1}$) and their
uncertainties.

In addition, following \citet{SingEtal2019ajWASP121bTransmissionNUV},
we also considered the Engineering Data Processing System ``jitter'' files,
which contain 28 different engineering measurements of {\HST}'s
Pointing Control System for each exposure during an observation.

\section{Data Analysis}
\label{sec:analysis}

The core of our analysis is based on the methodology of
\citet{CubillosEtal2020ajHD209458bNUV}.  We separated our light-curve
analysis into two main steps.  First we constructed ``white''
lightcurves by integrating the flux per transit visit and per echelle
order to boost the astrophysical and instrumental signals, from which
we characterized the instrumental systematics.  Then, we divided out
the systematics model from the raw data, and combined the
systematics-corrected data from all visits into spectral light curves
binned at different resolving powers, from which we extracted the
transmission spectra of the astrophysical signal.

\subsection{Analysis of Individual Visits}
\label{sec:white}

We characterized the instrumental systematics to detrend them
from the astrophysical signal and remove them from further analyses.
To increase the S/N of the data, we integrated
the flux over each echelle spectral order.  The implicit assumption is
that the instrumental systematics behave similarly over the integrated
range \citep{KreidbergEtal2014natCloudsGJ1214b}, in this case, over
each echelle order.

The analysis begins by masking bad and overly noisy data values.
Following the standard practice for exoplanet {\HST}/STIS time-series
analyses, we discarded the first orbit from each visit and the first
frame from each orbit, since they show significantly stronger
instrumental systematics \citep[see,
  e.g.,][]{SingEtal2011mnrasHD189733b_STIS}.  Then, we used the
overlap between echelle orders to discard data points where the
fluxes differ from each other by more than $3\sigma$.  Finally, we
manually inspected the edges of the echelle spectra to exclude data
points with abnormally low fluxes (typically, a handful of data points
at the ends of the echelle orders).
We then produced raw ``white'' light curves by summing the flux over
each spectral order and propagating the uncertainties accordingly.

We fit the raw light curves with parametric transit and systematics
models at each echelle order (denoted by their wavelength $\lambda$)
as a function of time ($t$), {\HST} orbital phase ($\phi$), and jitter
vector ($j_i$):
\begin{equation}
F_\lambda(t) = T_{\lambda}(t)\;S_{\lambda}(t, \phi, j_i),
\end{equation}
where $T_{\lambda}(t)$ is a
\citet{MandelAgol2002apjLightcurves} transit model and $S_{\lambda}(t, \phi,
j_i)$ is a model of the instrumental systematics.  To obtain
statistically robust parameter estimations, we computed best-fitting
parameters and uncertainties from a Levenberg-Marquardt optimization
and a Markov-chain Monte Carlo (MCMC) sampling, respectively.  For
this matter, we employed the open-source {\mcc} statistical
package\footnote{\href{https://mc3.readthedocs.io/}
  {https://mc3.readthedocs.io/}} \citep{CubillosEtal2017apjRednoise},
which implements the Snooker Differential-evolution MCMC algorithm of
\citet[][]{terBraak2008SnookerDEMC}.

The transit model parameterizes the astrophysical signal via the
orbital parameters, the planet-to-star radius ratios
{\rprs}($\lambda$), the out-of-transit stellar fluxes
$F\sb{\rm s}(\lambda)$, and limb-darkening coefficients
$C_i(\lambda)$.  For the orbital parameters we adopted previously
measured values for the orbital period
$P = 2.21857567 \pm 1.5\times10^{-8}$ day, inclination
$i = 85{\fdg}71$, and semi-major-axis to stellar radius ratio
$a/\rstar = 8.84 \pm 0.27$ \citep{StassunEtal2017ajGaiaRadiiMasses},
which we kept fixed during the fitting and MCMC process.  We fit
the transit mid-time epoch, applying a Gaussian prior according to
the measured optical value \citep[$T_0=53955.025551 \pm 9\times10^{-6}$ MJD,][]{BonomoEtal2017aaRVmasses}, propagating uncertainties according 
to the epoch of the {\HST} observations.

For the limb-darkening parameters, we computed theoretical
coefficients based on the stellar properties, since the low S/N ratios
and gaps in the {\HST} observations do not allow for a good
estimation.  For this calculation we used the open-source routines of
\citet{EspinozaJordan2015mnrasLimbDarkeningI} to compute
limb-darkening coefficients over each wavelength range based on the
PHOENIX stellar model \citep{HusserEtal2013aaPHOENIXstellarModels}
that best matches the properties of {\oneeightnine}: effective
temperature $T_{\rm eff}=5052$~K, surface gravity $\log g=4.49$, and
metallicity ${\rm [M/H]} = -0.02$ \citep{StassunEtal2017ajGaiaRadiiMasses}.
We note that, in absence of any better known guess, the adopted
limb-darkening coefficents correspond to the photosphere of the star.
If a significant fraction of the stellar flux comes
from its chromosphere and transition region, as suggested by the stellar 
line emission, then the used limb-darkening coefficients may not be
as representative as expected.

The {\HST}/STIS time-series observations exhibit a number of well
documented instrumental systematics: (1) periodic variations in phase
with the {\HST} orbital period that are suspected to originate from the
telescope's thermal cycle \citep{BrownEtal2001apjHD209458bHSTstis},
and (2) visit-long variations that might be attributed to stellar
activity \citep[e.g.,][]{WakefordEtal2016apjHSTsystematics,
  SingEtal2019ajWASP121bTransmissionNUV}.
Following the example of previous studies, we model these systematics
considering a family of polynomials in time ($t$) and {\HST}
orbital phase.
Furthermore, \citet{SingEtal2019ajWASP121bTransmissionNUV} showed that
additional data from {\HST}'s Pointing Control System (so called
``jitter'' data) can help decorrelate instrumental
systematics from the astrophysical signal.  For each science exposure,
the jitter data describe the {\HST}'s fine guidance sensor (FGS)
coordinates, FGS jitter, right ascension, declination, position angle,
latitude, longitude, angle between {\HST} zenith and target,
Earth-limb angle, terminator angle, and magnetic field.
{\HST} stores the jitter measurements into time-tag ancillary
engineering files, recorded at a high temporal resolution.
For each science exposure we computed a value for each of the jitter
parameters by calculating the median of the jitter values during the
exposure.
Therefore, our systematics model consists of a polynomial expression
up to a quadratic degree in time, up to a quartic degree in {\HST}
phase, and a up to a linear degree in one of the jitter vectors:
\begin{align}
\nonumber
& S_{\lambda}(t, \phi, j_i) =  1 +\ a_{0}(t-t_{0}) + a_{1}(t-t_{0})^{2}  \\
\nonumber
& \hspace{1.0cm} + b_{0}(\phi-\phi_{0}) + b_{1}(\phi-\phi_{0})^{2} 
  +\ b_{2}(\phi-\phi_{0})^{3} + b_{3}(\phi-\phi_{0})^{4} \\
& \hspace{1.0cm} + c_0(j_i-\langle{j_i}\rangle),
\label{eq:systematics}
\end{align}
where $a_{k}$, $b_{k}$, and $c_{k}$ are the polynomial coefficients to
be fit.  Note that we considered an individual set of fitting
parameters for each echelle order (we dropped the $\lambda$-dependence
of these coefficients for clarity).  Regarding the jitter vectors
$j_i$, we considered each of the 21 vectors listed in Table 1
of \citet[][]{SingEtal2019ajWASP121bTransmissionNUV}.  $t\sb{0}$ and
$\phi\sb{0}$ are reference values for the time and phase, which are fixed
at the transit mid-time $t\sb{0}=T\sb{0}$ and at the {\HST} mid-phase
$\phi\sb{0}=0.2$, respectively. $\langle{j_i}\rangle$ denotes the mean
value of the jitter vector $j_i$ along the visit.

\begin{figure*}
\centering
\includegraphics[width=0.95\linewidth]{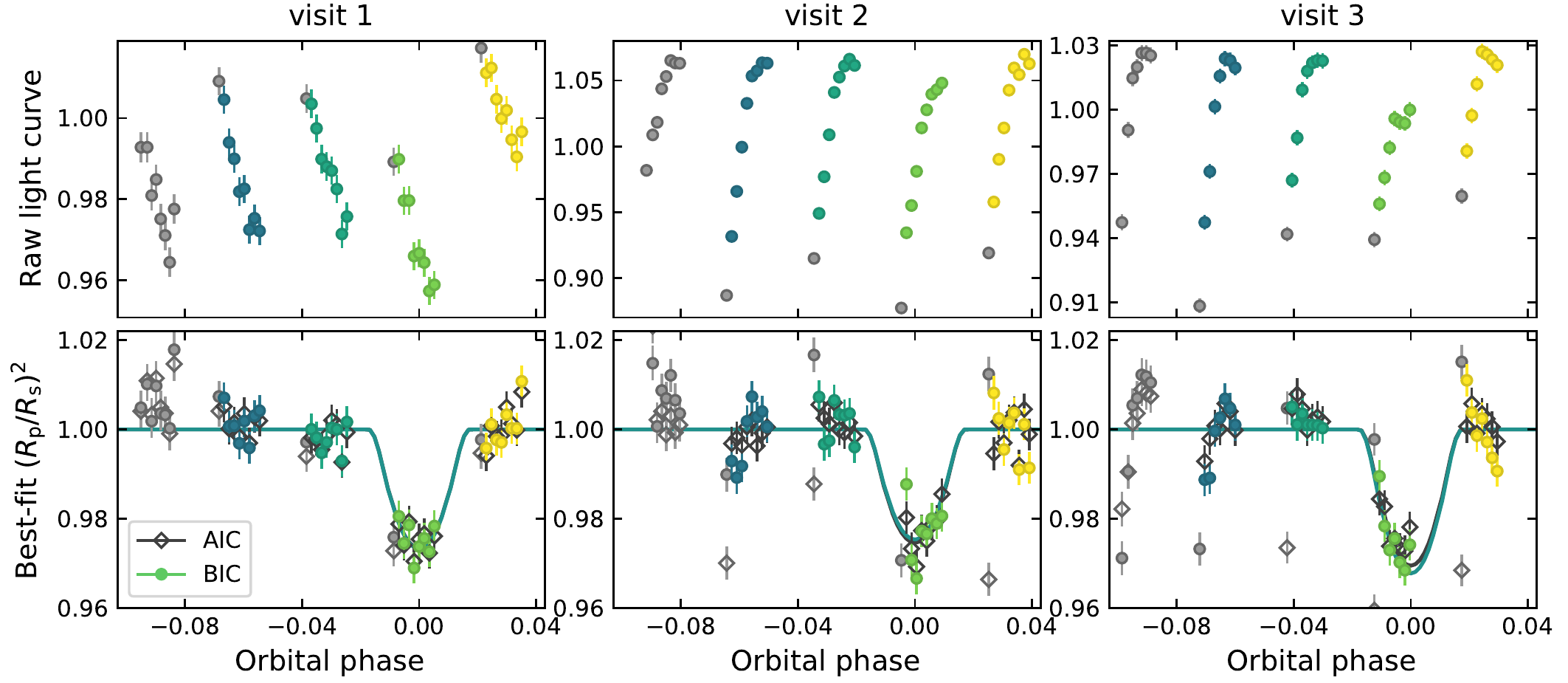}
\caption{Sample transmission light-curve fitting of the {\HST}/STIS
  observations of {\oneeightnineb} (fourth echelle order,
  $\bar \lambda=2917 \AA$).  The top row shows the raw light curves
  (one column for each of the three visits).  The markers with error
  bars denote the fluxes and their uncertainties integrated over the
  echelle order for each frame.  The markers are color-coded according
  to the {\HST} orbit, where the gray colors denote the frames
  discarded from the light-curve fitting.  The bottom rows show the
  systematics-corrected light curves corresponding to the best-BIC
  (colored circles) and best-AICc statistics (empty black diamonds),
  including the jitter decorrelation.  The solid lines denote the
  astrophysical model of the best-BIC (colored lines) and best-AICc
  models (black lines).}
\label{fig:white_light_curves}
\end{figure*}

\subsubsection{Systematics Model Selection}

Given our limited understanding of the systematics
affecting {\HST} observations, we determined the optimal functional
form for the systematics model $S_\lambda$ by comparing fits 
considering all possible combinations of polynomial orders in $t$ and
$\phi$ up to the degrees shown in Eq. (\ref{eq:systematics}), and
each of the 21 jitter vectors, and fits with no jitter vectors.
To this end we employed a Bayesian model selection approach
\citep[see, e.g.,][]{Trotta2007mnrasBayesianModelSelection}, where one
compares the posterior probability of two competing
models---given the data---by computing the ratio of their Bayesian
evidences (also known as Bayes factor).
The main problem of this approach is that the evidence of a model is
not readily calculable.  Numerically, calculating the evidence entails
evaluating the model likelihood throughout its parameter space (e.g.,
via a nested-sampling algorithm).  To avoid the computational cost of
this task, we adopt a simpler model selection approach where we
approximate the evidence ratio via the Bayesian Information Criterion
\citep[BIC,][]{Liddle2007mnrasBIC}:
\begin{equation}
{\rm BIC} = \chi^{2}_{\rm min} + k\log N,
\end{equation}
where $\chi^{2}_{\rm min}$ is the goodness-of-fit parameter
corresponding to the maximum likelihood for the given model, $k$ is
the number of free parameters, and $N$ is the number of data points.
The Bayes ratio is then simply calculated from the difference in BIC
values between the competing models $B = \exp\{-({\rm BIC}_1 - {\rm
  BIC}_2)/2\}$ \citep{Raftery1995BIC}, which only requires to find the
maximum likelihood for the given models.  A Bayes factor $B>1$ favors
model 1 over model 2, or alternatively, the favored model minimizes
BIC.

Certainly, these approximations are based on various assumptions, which may not
be fulfilled to different extents, such as independent and identically
distributed data or (nearly) Gaussian posterior distributions or that the forward model is correct, for example.
Therefore, we also considered a second model-selection criterion based
on information theory, that is the Akaike Information Criterion
corrected for small sample sizes \citep[AICc,][]{Liddle2007mnrasBIC}:
\begin{equation}
{\rm AICc} = \chi^{2}_{\rm min} + \frac{2kN}{N-k-1}.
\end{equation}
Just like before, the favored model minimizes AICc.

\begin{table}[t]
\centering
\caption{BIC/AICc Model Comparison}
\label{table:statistics}
\begin{tabular} {@{\extracolsep{-0.1cm}} lcccl}
\hline
\hline
Systematics model & BIC   & AICc         & $\chi^2_{\rm red}$ & Comments \\
{($t$-deg, $\phi$-deg, jitter)} &  &  &   \\
\hline
\multicolumn{1}{l}{\bf Visit 1} \\
(1, 1)      & 1362.2  &  961.6  & 1.16  & best BIC \\
(1, 1, Lat) & 1434.5  &  944.6  & 1.08  & best AICc \\
(1, 2)      & 1441.5  &  951.6  & 1.09  & best AICc no-jit \\
\\
\multicolumn{1}{l}{\bf Visit 2} \\
(1, 1, Lat) &  1617.7  &  1127.8  & 1.37  & best BIC \\
(1, 3, Lat) &  1692.9  &  1039.7  & 1.09  & best AICc \\
(1, 2)      &  1699.0  &  1209.1  & 1.50  & best BIC no-jit \\
(2, 2)      &  1740.3  &  1166.1  & 1.38  & best AICc no-jit \\
\\
\multicolumn{1}{l}{\bf Visit 3} \\
(1, 1, LOS Zenith) &  1665.0   &  1198.2  & 1.73  & best BIC \\
(1, 3, LOS Zenith) &  1704.7   &  1090.8  & 1.36  & best AICc \\
(1, 2)             &  1798.4   &  1331.6  & 1.98  & best BIC no-jit \\
(2, 2)             &  1826.6   &  1282.8  & 1.84  & best AICc no-jit \\
\hline
\end{tabular}
\tablefoot{The first two terms in the systematics model column show
  the polynomial degree for time and orbital phase, the third term
  shows the name of the jitter vector (which is always modeled as a
  linear polynomial).  The absence of a third term indicates that no
  jitter vector was included.}
\end{table}

\subsubsection{Model Selection Applied to the {\HST} Observations}

For the {\oneeightnineb} data we treated each {\HST} visit
independently of each other, where for each visit we fit
simultaneously the astrophysical and systematics parameters.  Also,
for each visit we fit all 23 echelle orders simultaneously, jointly
fitting the transit mid-time epoch and adopting the same configuration
of the systematics model for all orders.  Table \ref{table:statistics}
reports the BIC and AICc of the global fit to all orders for the
favored systematics model (with and without jitter decorrelation).
Figure \ref{fig:white_light_curves} shows an example of the light-curve
fits for a selected echelle order.  

As seen from the raw light curves (top row of
Fig.~\ref{fig:white_light_curves}), the systematics are characterized
by visit-long and {\HST}-orbit-long variations.  Overall, the second
visit shows stronger systematics than the first visit, and the third
visit shows stronger systematics than the second, which is reflected
in the best-fitting BIC, AICc, and reduced $\chi^{2}$ values
($\chi^{2}_{\rm red}$).  In particular, the second and third visit
systematics vary on the order of 10\% of the raw flux.  If we only
included the more ``traditional'' polynomials in $t$ and $\phi$ in the
fit, we found clear systematic residual trends, which vary in strength
for each echelle order.  The relatively new practice of jitter
decorrelation introduced by
\citet{SingEtal2019ajWASP121bTransmissionNUV} significantly improves
the fit for the second and third visits, minimizing both information
criteria statistics and residual systematics (bottom row of
Fig.~\ref{fig:white_light_curves}).
In these fits, the AICc imparts a weaker penalty per fitting parameter
than the BIC, thus favoring more complex models, which notably
improves the $\chi^{2}_{\rm red}$ values.  However, although the BIC
and AICc statistics favor different models, the two criteria usually
produced consistent transit depths. All following analyses presented 
here will focus on the best AICc analysis including jitter
decorrelation.

\begin{figure}[t]
\centering
\includegraphics[width=\linewidth]{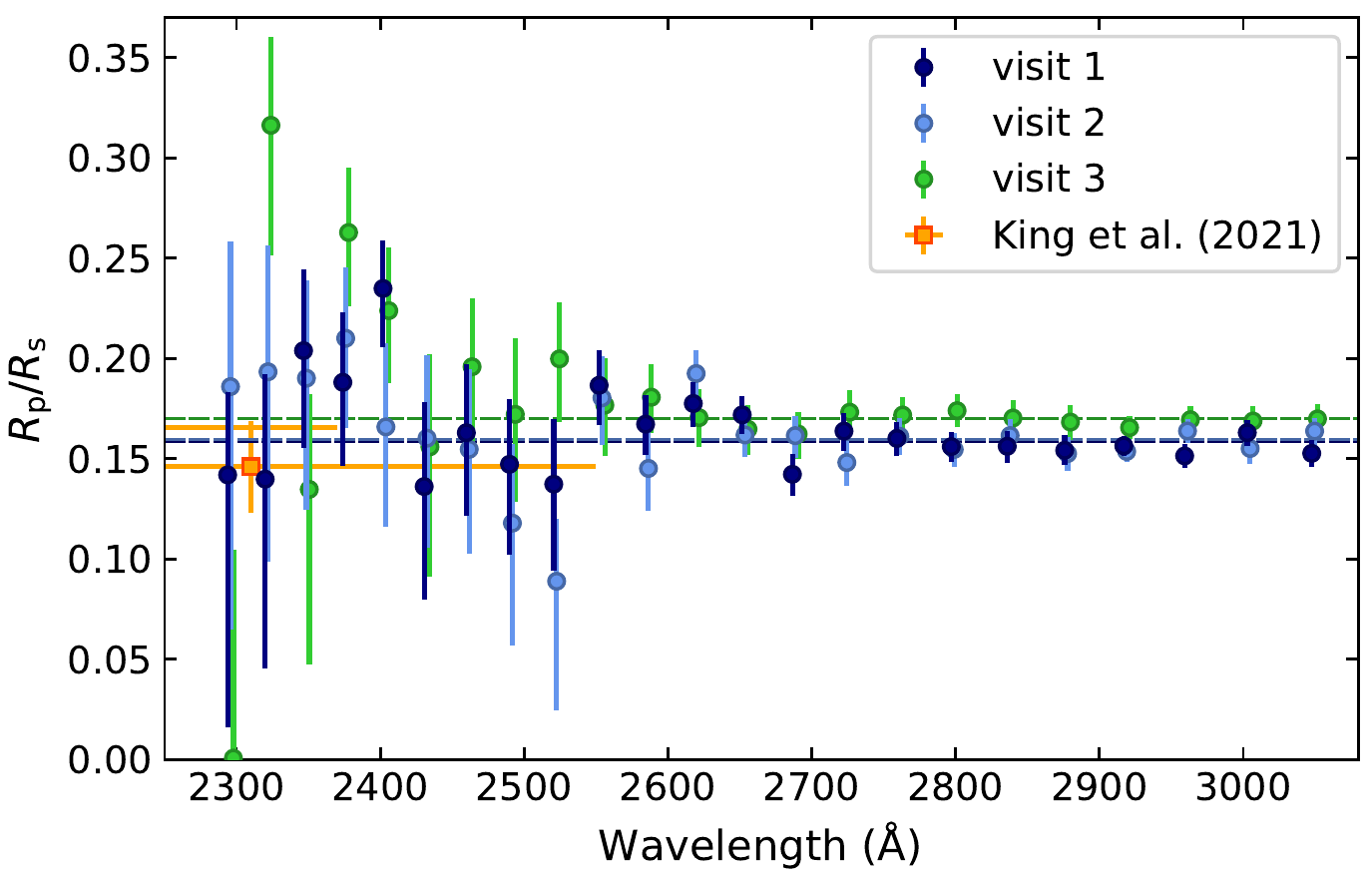}
\caption{{\HST}/STIS NUV transmission spectrum of {\oneeightnineb}
  (individual visits). The circle markers with error bars denote the
  transmission planet-to-star radius ratios from each echelle order
  and visit (see legend), corresponding to the best-AICc fits.  The
  colored dashed lines denote the mean value for each visit.
  The square marker denotes the {\em XMM-Newton}/UVM2
  photometric measurement by
  \citet{KingEtal2021mnrasNUVxmmHD189733b}.
}
\label{fig:white_transmission}
\end{figure}

Figure \ref{fig:white_transmission} shows the resulting transmission
spectra according to the best-AICc models for each visit and at each
echelle order.
One clear feature seen in the transmission spectra is an offset
between the third visit and the first two.  This offset is consistent
with the signature of unocculted stellar spots \citep[see,
e.g.,][]{RackhamEtal2018apjStellarHeterogeneityI,
  ZellemEtal2017apjTransitStellarActivity}, which is a likely cause
given the well known active nature of {\oneeightnine}.
For each visit, we found that the transit radius ratios vary closely
proportional to the square root of the bolometric fluxes of the observations
($\rprs \propto \fstar^{-1/2}$), which is a direct consequence of
observations arising from a heterogeneous stellar photosphere
including spots \citep[][]{RackhamEtal2018apjStellarHeterogeneityI}.
The weighted mean of the {\rprs} values for each individual
visit are 0.1585, 0.1595, and 0.1701; whereas
the relative bolometric fluxes between visits are
$F_{\rm s}^{\rm v1}/F_{\rm s}^{\rm v2} = 1.02$ and
$F_{\rm s}^{\rm v1}/F_{\rm s}^{\rm v3} = 1.16$.  Later, the combined
analysis in Section \ref{sec:coarse} will show an independent
confirmation that we are observing stellar spot contamination in the
data.

Once corrected for stellar activity, the analysis of each of the three
visits produces a consistent transmission spectrum across the NUV.  We
obtained a mostly flat transmission spectrum between 2300 {\AA} and
3100 {\AA}, with a planet-to-star radius ratio of $\sim$ 0.16, which
becomes significantly noisier at wavelengths shorter than 2550 {\AA},
due to the much lower stellar flux.
These results are also consistent with the previous NUV observation by
\citet{KingEtal2021mnrasNUVxmmHD189733b}, which measured the transit
depth of {\oneeightnineb} in the UVM2 photometric band of the {\em
  XMM-Newton} Space Observatory (effective wavelength 2310 {\AA},
width 480 {\AA}).

As a reference, the estimated altitude of the Roche lobe radius as probed during transit is $R_{L1'} = 0.44~\rstar$ (i.e., perpendicular to the star--planet axis).  This value is approximated by 2/3 the distance to the L1
Lagrange point \citep[see, e.g.,][]{VidalMadjarEtal2008apjHD209458bEvaporation}.
All transit depths at each echelle order are well under the Roche lobe radius.

\subsection{Analysis of Combined Visits}
\label{sec:spectra}

To search for the narrow metal line-absorption features in the
transmission spectrum, we need to analyze the observations at a
high resolving power, finding a compromise between sufficient S/N per
wavelength channel and diluting the narrow absorption features.
Therefore, we treated the data at multiple spectral resolving powers
ranging from $R=10-70$ (coarse sampling analysis),
and then at a $R=4700$ (fine sampling analysis).
We proceeded in the same fashion as in
\citet{CubillosEtal2020ajHD209458bNUV}, that is, calibrate the
wavelength solution to ensure that the spectra are well aligned, use
the modeling results from the previous analysis to divide out the
instrumental systematics, combine the data from all three visits into
a single spectrum, and fit the transit depths at each bin.  All future
analyses use the best-AICc dataset since it attained
$\chi^2_{\rm red}$ values closer to one.

\subsubsection{Wavelength Calibration}

We used a cross-correlation procedure to calibrate the wavelength
solution of each echelle order in each frame and visit.
First, we created a master stellar template (enhancing the S/N) by
co-adding the spectra from all frames within an echelle order and
visit.
We then Doppler-shifted the individual-frame spectra and
cross-correlated them with their template, searching for the 
relative velocity with respect to the template given by the shift that
maximized the correlation function.
Once we obtained the wavelength correction of the individual frames
within an echelle order, we repeated the procedure between the
template spectra of each visit to find the relative wavelength
correction between all spectra.  Finally, we compared our wavelength
solution with a theoretical model spectrum of {\oneeightnine}
\citep{ShulyakEtal2004aaLBLstellarModels} to obtain an absolute
wavelength solution at a rest reference frame.
Once calibrated, we Doppler-shifted all frames within a given echelle
order to a common wavelength solution using a cubic spline
interpolation.  To minimize the impact of the interpolation, we
selected a reference point that minimized the shifts of the set of
frames.
For a given visit, we found offsets of less than half a pixel between
the frames, and in general the wavelength solution for each echelle
order follows a similar trend along the visit, with no outliers.

\subsubsection{Divide-white Spectral Extraction}

To extract a high-resolution transmission spectrum of
{\oneeightnineb}, we followed the ``divide-white'' spectral analysis
by \citet{KreidbergEtal2014natCloudsGJ1214b}, as described in
{\citet{CubillosEtal2020ajHD209458bNUV}}.
In this approach, we first constructed a non-parametric model of
the instrumental systematics by dividing the white-light curves (for
each echelle order and visit) by their best-fit astrophysical transit
model (see Sect.~\ref{sec:white}).  Assuming that the instrumental
systematics vary weakly with wavelength, each raw spectral data point
is divided by the systematics model corresponding to their orbital
phase, echelle order, and visit.  Data uncertainties are properly
accounted for, i.e., using the white-light model's posterior
distribution to estimate the uncertainties systematics model and
propagating the errors.

In a second step, we split the data into constant resolving-power
wavelength bins,
where for each spectral channel we co-added all flux contributions from
all echelle orders, and propagated the errors accordingly.

To analyze the spectral light curves we fit a
\citet{MandelAgol2002apjLightcurves} transit model to the
systematics-corrected spectral light-curves in an MCMC run.  As
previously, we fixed the orbital $a/R\sb{\rm s}$, $\cos(i)$, and $P$
parameters, and applied a Gaussian prior for the mid-transit epoch
based on the white-light fitting results.
We also computed theoretical limb-darkening curves at each spectral
channel, which we kept fixed during the fit.  At each wavelength bin
we fit a unique transit depth common to all three visits, but we fit a
separate out-of-transit flux value for each visit.  Additionally, we
incorporated a stellar-spot correction model that we detail in the
following section.

\subsubsection{Stellar Spot Correction}
\label{sec:spots}

To account for the spectral transit-depth variations due to unocculted
stellar spots, we modeled the stellar flux as a linear combination of
a nominal photospheric region and a spotted region.
We assumed that a fraction {\fspot} of the observed photosphere is
covered by spots with a characteristic temperature {\tspot}, whereas
the rest of the photosphere has the nominal effective temperature of
the star $\tstar = 5052$~K (with $\tstar \geq \tspot$).  We used the
PHOENIX library of high-resolution synthetic stellar spectra
\citep{HusserEtal2013aaPHOENIXstellarModels} to compute the flux of
the nominal stellar photosphere $\fstar(\tstar)$ and the spotted
region $\Fspot(\tspot)$.  This leads to the following correction term
for the observed transit depth $d_{\rm obs}(\lambda)$:
\begin{equation}
d_{\rm obs}(\lambda) = \frac{d(\lambda)}
    {1-\fspot\left[1-{\Fspot(\lambda)}/{\fstar(\lambda)}\right]}
   \equiv d(\lambda)\ \epsilon(\lambda),
\end{equation}
where $d(\lambda)$ is the true transit depth and $\epsilon(\lambda)$
is the contamination spectrum due to stellar hetereogeneities
\citep[as defined in][]{RackhamEtal2018apjStellarHeterogeneityI}.  To
evaluate the correction factor at any given {\tspot}, we interpolated
in temperature between all available PHOENIX models (ranging from
2300~K to 5100~K) at a fixed metallicity ([M/H]=0.0) and surface
gravity ($\log(g)=4.5$).

\begin{figure*}[t]
\centering
\includegraphics[width=0.95\linewidth]{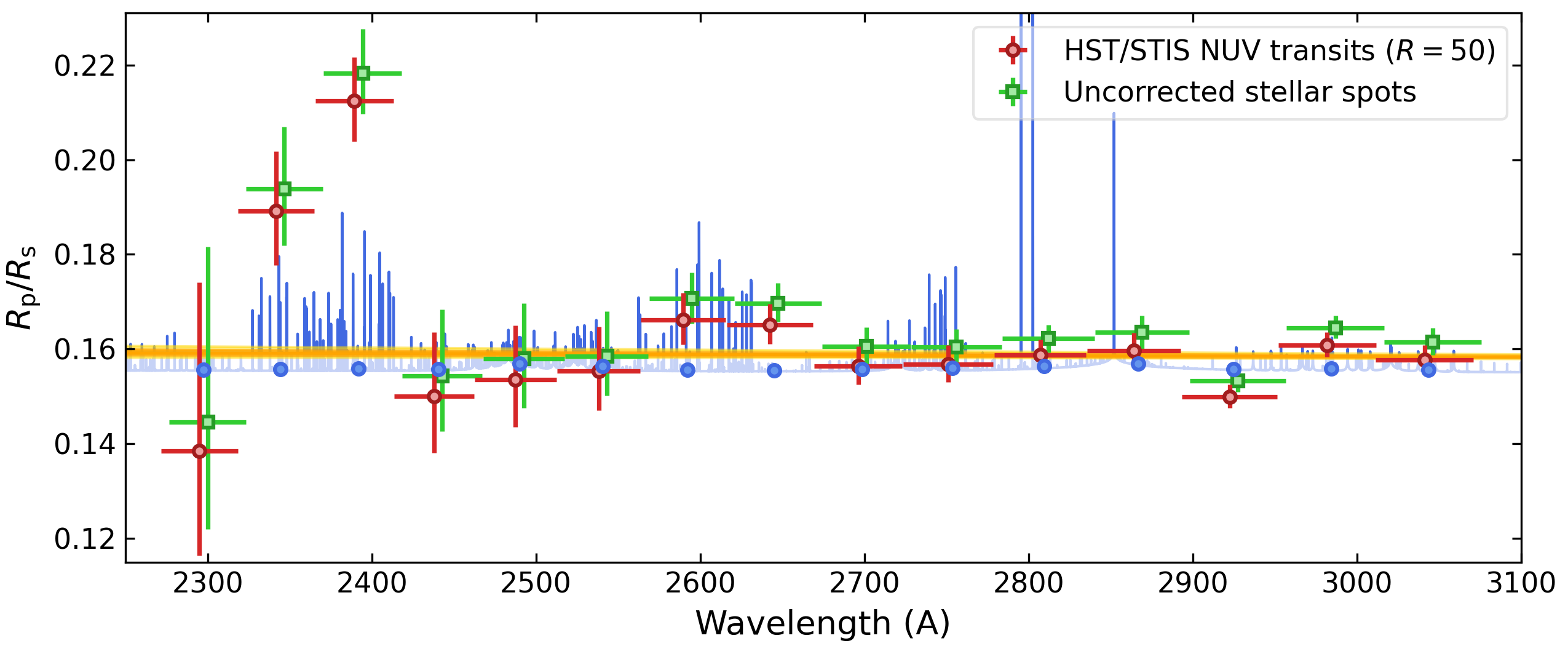}
\caption{NUV transmission spectrum of {\oneeightnineb} from the
  combined {\HST}/STIS observations.  The red markers with error bars
  denote the systematics- and stellar-spot corrected transmission
  spectrum, their 1$\sigma$ uncertainties, and the span of the
  spectral bins (the data has been binned in wavelength over a
  resolving power of R\,=\,50).  The green markers show our results when
  neglecting the stellar spot contamination in the analysis (markers
  have been slightly shifted in wavelength for visibility). The blue
  spectrum shows a theoretical model of the planet's upper atmosphere
  (see Section \ref{sec:upper_model}) and the blue dots show the model
  integrated over the spectral bins.  The orange curve shows an
  extrapolation from a fit to the optical-IR transmission
  spectrum of the planet (see Section \ref{sec:lower_model}).}
\label{fig:coarse_transmission}
\end{figure*}

The stellar-spot correction model thus contains two free parameters
for each visit ({\fspot} and {\tspot}), which we fit simultaneously
along with the other astrophysical parameters during the MCMC.  We
considered a uniform prior for the {\fspot} values between 0.0 and 0.5
(it would be unlikely to find unocculted spot covering fractions
greater than 50\%).  For the spot temperature we considered a
logarithmic-uniform prior in $\Delta T_{\rm spot} = \tstar-\tspot$ because the
spectral variation of the spot correction is more pronounced at small
$\Delta T_{\rm spot}$ values \citep{McCulloughEtal2014apjHD189733bHST}.  When
$\Delta T_{\rm spot} \gtrsim 1500$~K the correction becomes nearly achromatic.
Regardless of the parameterization, the main conclusions of our
analysis do not change when we consider a uniform prior in $\tspot$.

\section{Planetary Atmospheric Signatures}
\label{sec:spectra}

\subsection{Coarse-binning Spectra}
\label{sec:coarse}

To boost the characterization of the astrophysical signal, we started
with the analysis of the combined data binning over a relatively coarse
wavelength sampling.  Figure \ref{fig:coarse_transmission} shows the
NUV transmission spectrum of {\oneeightnineb} sampled at a resolving
power of ${\rm R} = 50$. This resolving power resulted in the
best compromise between having a good S/N per bin and
having sufficient spectral resolution to identify variability with
wavelength.  Appendix \ref{sec:coarse_lightcurves} shows the
corresponding systematics- and stellar-spot corrected lightcurves for
this analysis.
At this resolving power, we obtained a relatively flat transmission
spectrum with the exception of two regions where the transit depth
increases (by more than $\sim$1$\sigma$ relative to the surrounding
wavelengths) located at $\lambda\sim2350$~{\AA} and
$\lambda\sim2600$~{\AA}.

For completeness, we also analyzed the data at other similar resolving
powers of R = $\sim$$10, 30,
70$ finding consistent results.  Likewise, analyses with the bins
shifted by half a bin-width returned consistent results as well.

\begin{figure}[ht]
\centering
\includegraphics[width=\linewidth]{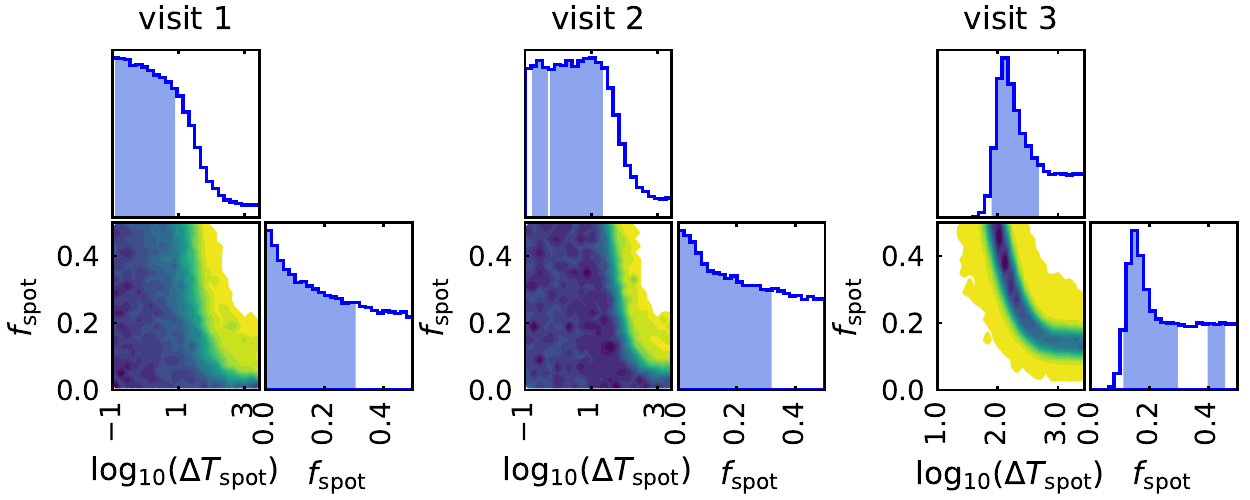}
\includegraphics[width=\linewidth]{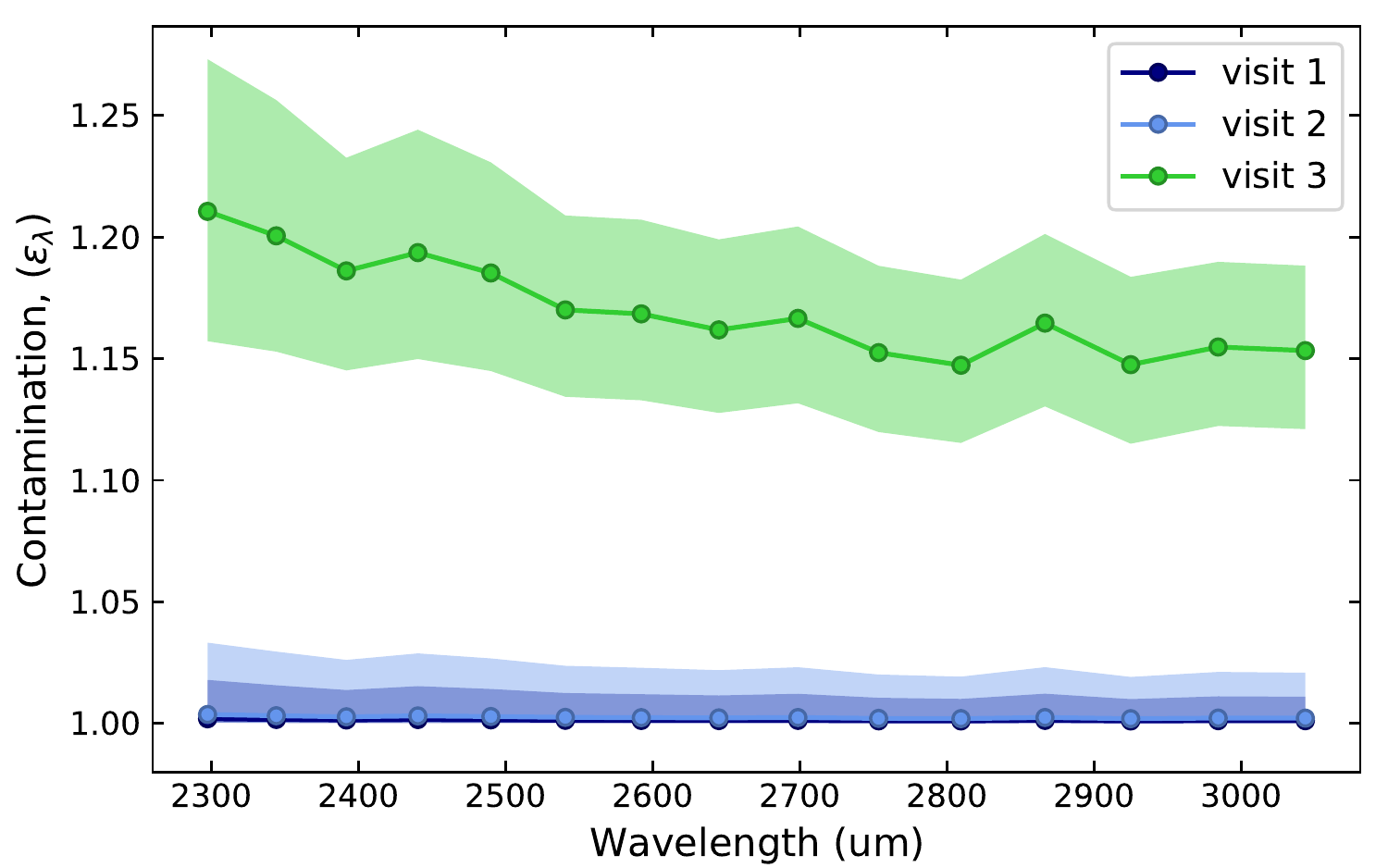}
\caption{Stellar spot contamination. The top panels show the posterior
  distribution of the spot parameters derived from the combined-data
  analysis.  The blue shaded areas denote the 1$\sigma$ highest
  posterior density credible region.  Visits 1 and 2 show a negligible
  spot contamination with either a small spot covering fraction or a
  negligible difference between the spot and photospheric temperatures.
  Visit 3 in contrast shows a non-negligible contamination of
  $\epsilon(\lambda)\sim$1.15--1.2, with a strong correlation
  between the spot covering fraction and spot temperature.  The bottom
  panel shows the respective contamination spectra.  The colored
  markers and vertical spans show the median and 1$\sigma$ uncertainty
  of the contamination, respectively.}
\label{fig:contamination}
\end{figure}

\begin{figure*}[t]
\centering
\includegraphics[width=\linewidth]{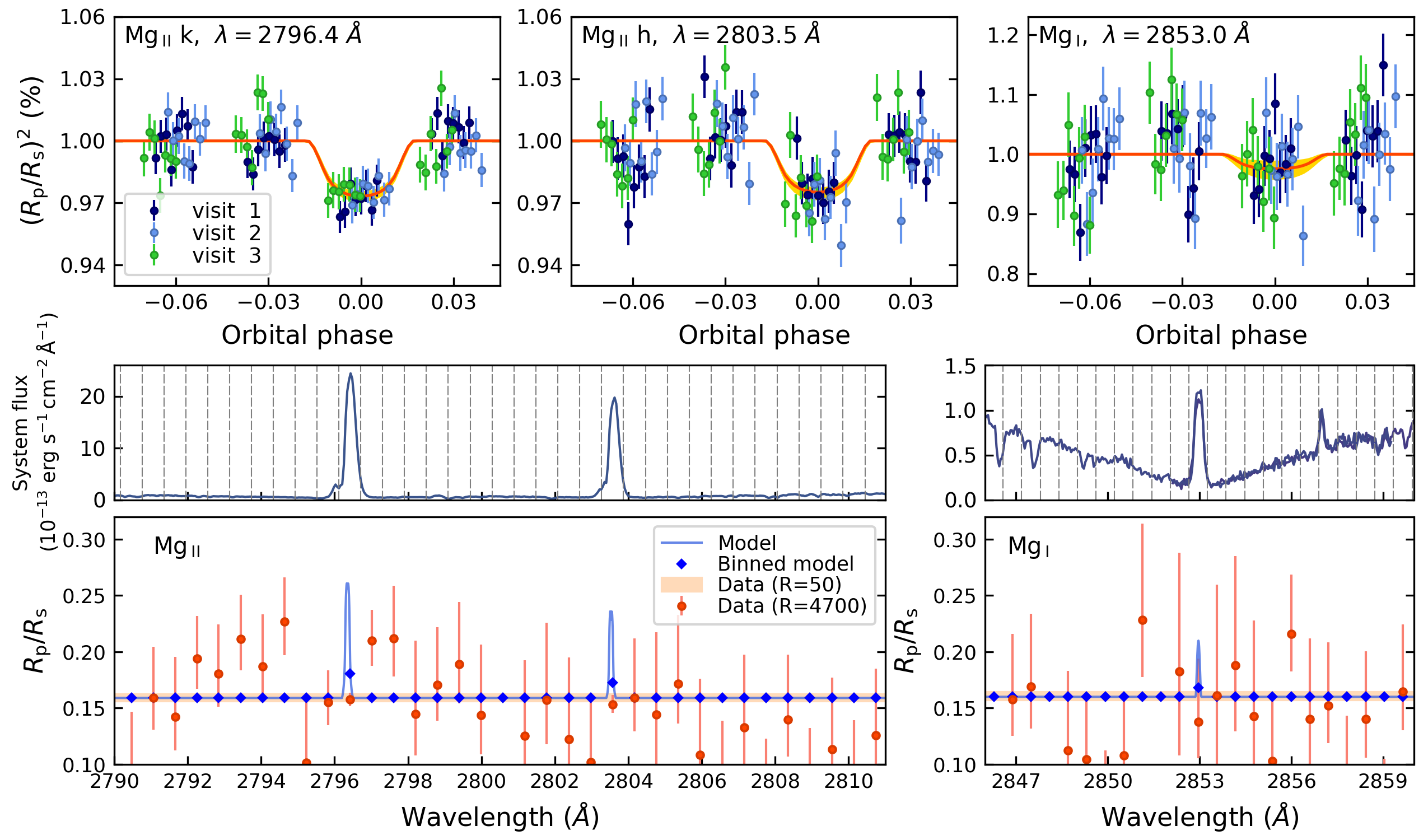}
\caption{Transit light curves (top) and spectrum (bottom) of
  {\oneeightnineb} of the combined {\HST}/STIS transits at the
  location of the magnesium resonance lines, and binned at a resolving
  power of $R=4700$ (i.e., roughly the width of the magnesium lines).
  The top panels show the systematics-corrected light curves for the
  spectral channels centered at the core of the {\mgii} and {\mgi}
  lines (see labels). The solid red curve and orange area denote the
  best-fitting transit depths and span of their 1$\sigma$
  uncertainties.  The middle panels show the system flux around the
  magnesium lines, with the vertical dashed lines marking the edges of
  the spectral bins.  The bottom panels show the transit
  planet-to-star radius ratio for each of the bins around the
  magnesium resonance lines (red markers with 1$\sigma$ error bars).
  The blue curve shows the upper-atmosphere model (the diamond markers
  show the model binned at $R=4700$). The peach-colored area shows the
  span of the low-resolution transit radius ratio ($R=50$).}
\label{fig:magnesium}
\end{figure*}

The stellar spot contamination was well characterized, as shown in
Figure \ref{fig:contamination}.  The first two visits show a
negligible spot contamination
($\epsilon<1.05$), with the models finding either a small spot covering
fraction or a spot temperature close to the nominal stellar
effective temperature ($\Delta T_{\rm spot} \lesssim10$).
In contrast, the third visit shows a clearly significant spot
contamination ranging from
$\epsilon(\lambda)\approx1.20$ to 1.15 from short to long
wavelengths.  The contamination shows a weak variation with wavelength,
which is not larger than the
1$\sigma$ credible intervals of the constraints.  Accordingly, we
constrained the spot covering fraction to $f_{\rm
  spot}>0.11$ and the spot temperature to $100~{\rm K} \lesssim \Delta
T_{\rm spot} \lesssim 500~{\rm
  K}$, values which are strongly correlated (see posterior pair-wise
distribution in Fig.\ \ref{fig:contamination}).  Analyzing the
combined data without a stellar-spot model produced slightly larger
transit depths (Fig.\ \ref{fig:coarse_transmission}); however, the
shape of the spectrum remains consistent with that of the
spot-corrected analysis, which is consistent with the nearly
wavelength-independent stellar spot correction.

All in all, the ratio between contamination spectra of the visits
(e.g., $\epsilon^{\rm v3}/\epsilon^{\rm v1} = 0.15-0.2$) matches
precisely the ratio between bolometric fluxes (e.g.,
$F_{\rm s}^{\rm v1}/F_{\rm s}^{\rm v3} = 0.16$), which reinforces the
hypothesis that the variability between visits is caused by a varying
stellar spot coverage at different epochs.

\subsection{Fine-binning Spectra: Magnesium Resonance Lines}
\label{sec:magnesium}

We attempted to analyze the data binning at much higher resolving
power such that we can capture absorption of individual metallic
lines, however, the high-resolution lightcurves were largely dominated
by noise, and not much could be concluded from the resulting spectra.
The only region where we were able to place significant constraints at
high resolution is at the core of the magnesium resonance lines, which
have fluxes nearly ten times larger than at other wavelengths, and
thus a much better S/N.  For this, we selected a bin resolving power
of $R=4700$ (65 {\kms} wide bins), which encapsulates the core of the
{\mgi} and {\mgii} resonance lines.

Figure \ref{fig:magnesium} shows the high-resolution light curves
centered at the {\mgi} and {\mgii} resonance (top), and the stellar
spectra (middle) and transit radius-ratio spectra (bottom) around the
magnesium lines.  For the {\mgii} bins, the light curves show a clear
transit signature (albeit with some residual noise for the {\mgii} h
line), and have S/Ns on the same order of magnitude as those of the
low-resolution analysis.  The light curve at the {\mgi} line, having
much lower flux, is significantly noisier than the {\mgii} lines.
For all three magnesium lines, the
transit depths do not show any significant excess above the
low-resolution continuum values ($R=50$ analysis),
and are consistent with the optical-IR transmission spectrum of {\oneeightnineb}.

For context, in the bottom panels of Fig.\ \ref{fig:magnesium} we also
show the theoretical upper-atmospheric model for this planet (see
Sect.\ \ref{sec:upper_model} for details).  While {\oneeightnineb}'s
escape rate is not high enough to generate magnesium absorption
signatures close to the the Roche-lobe altitude
($R_{L1'}/\rstar = 0.44$), the model predicts a noticeable excess
absorption above the continuum.  Thus, we used the binned model
estimate to place non-detection limits for the {\mgii} measurements.
We found that the observations sit 2--4$\sigma$ below the prediction
for a solar-abundance atmospheric model containing Mg in the upper atmosphere
(Table \ref{table:magnesium}).  Due to the higher noise at the {\mgi}
line, we cannot reject nor confirm the presence of {\mgi} absorption
from our observations.

{\renewcommand{\arraystretch}{1.5}
\begin{table}[h]
\centering
\caption{{\oneeightnineb} Transit Measurements at Magnesium Lines}
\label{table:magnesium}
\begin{tabular} {lcc@{}c}
\hline
\hline
Line & $\lambda_0$ ($\microns$) & $\rprs$  & Non-detection significance$^\dagger$ \\
\hline
{\mgii} k & 2796.4  &  $0.158^{+0.006}_{-0.006}$ & 3.7$\sigma$ \\
{\mgii} h & 2803.5  &  $0.153^{+0.008}_{-0.007}$ & 2.3$\sigma$ \\
\mgi      & 2853.0  &  $0.138^{+0.056}_{-0.032}$ & $<$1$\sigma$ \\
\hline
\end{tabular}
\begin{tablenotes}
  \item $^\dagger$ number of standard deviations below the expected {\rprs}
  from the solar-abundance theoretical model.
\end{tablenotes}
\end{table}
}

\section{The NUV-to-Infrared Transmission Spectrum of {\oneeightnineb}}
\label{sec:context}

For context, the He{\sc i} infrared metastable triplet and 
H{\sc i} Lyman-$\alpha$ transit observations of
{\oneeightnineb} can be interpreted in terms of the planet 
hosting a compact upper atmosphere;
extended, but not as much as that of other
exoplanets observed in the NUV, e.g., {\twoohnineb}.  Metals like
magnesium or iron may condense in the lower atmosphere and even if 
they make it to the upper
atmosphere, a lower escape rate would not allow their density profiles
to stretch out to high altitudes.  At longer wavelengths, the
optical-IR observations show a strong slope towards the blue,
which can be interpreted as absorption from high-altitude hazes.

To test these hypotheses and to get a better understanding of the
interaction between the upper and lower atmosphere of
{\oneeightnineb}, we compare the observations to theoretical models.
Since it is not possible to model all required upper- and
lower-atmosphere properties at once, we model individually the upper
and lower regions, and then study how they relate to each other.

\subsection{Upper-atmosphere Modeling}
\label{sec:upper_model}

To interpret the NUV transmission spectrum of {\oneeightnineb}, we
employed the physical model of the upper atmospheres
of close-in planets \citep{KoskinenEtal2013icarHD209458bMetalsEscapeI,
  KoskinenEtal2013icarHD209458bMetalsEscapeII,
  KoskinenEtal2022apjExtremePlanetsLoss}. Since the parts of the NUV
transmission spectrum that probe the upper atmosphere are dominated by
signatures of ionized metals {\feii} and {\mgii}
\citep{SingEtal2019ajWASP121bTransmissionNUV,
  CubillosEtal2020ajHD209458bNUV}, we added the heavy elements C, O,
N, Mg, Si, Fe, S, Ca, Na, and K with solar abundances to the model, in
addition to H and He, with related chemistry and physics adapted from
\citet{HuangEtal2015apjHD189733bUpperModeling}. We included Mg, Si,
Fe, Ca with first and second ionization states, due to the relatively
low ionization potential of the first ionization state, and the rest
of the heavy elements with their first ionization state. The electron
density is equal to the sum of the ion densities under the assumption
of quasineutrality. The lower boundary of the escape model is at 0.1
$\micro$bar where the temperature is sufficient to dissociate molecules
that are not included in the model.

We used the results from the photochemical model of
\citet{LavvasKoskinen2017apjAerosolProperties} to obtain the mean
molecular weight profile in the lower and middle atmosphere to set the
lower boundary altitude and temperature of the escape model. The
temperature profile in the lower and middle atmosphere was not
calculated self-consistently by the photochemical model and was
instead adapted from
\citet{MosesEtal2011apjDissequilibriumHD209nHD189b}. Here, we used the
MK profile from \citet{LavvasKoskinen2017apjAerosolProperties}. The
application of the photochemical model to HD189733b did not include Mg
and Fe that are detectable in the NUV. In order to crudely model the
full profiles of the absorption lines, we added Mg and Fe to the lower
and middle atmosphere, assuming solar abundances and thermal
ionization according to the Saha equation
\citep{Menou2012apjMagneticScaling}. This approximation does not
affect our conclusions because our focus in this section is on the
possible upper atmosphere signatures. 

Given the temperature profile assumed for {\oneeightnineb} here and assuming
equilibrium condensation, Mg, Fe, Si, and Ca are expected to condense
to form mineral clouds and rain out from the atmosphere
\citep[e.g.,][]{WakefordSing2015aaTransitCondensates}. We included Mg
and Fe in the escape model simply to explore if the modeled absorption
lines of Mg II and Fe II would be detectable in the NUV data. The rest
of the heavy elements were retained for consistency with the
no-condensation scenario. Although efficient mixing in the lower
atmosphere could interfere with mineral cloud formation and allow for
higher abundances of the relevant heavy elements in the upper
atmosphere \citep{SpiegelEtal2009apjTiO,
  KoskinenEtal2013icarHD209458bMetalsEscapeII}, we do not hold a view
on whether this is possible on {\oneeightnineb} or not at this point.

\begin{figure*}[t]
\centering
\includegraphics[width=\linewidth]{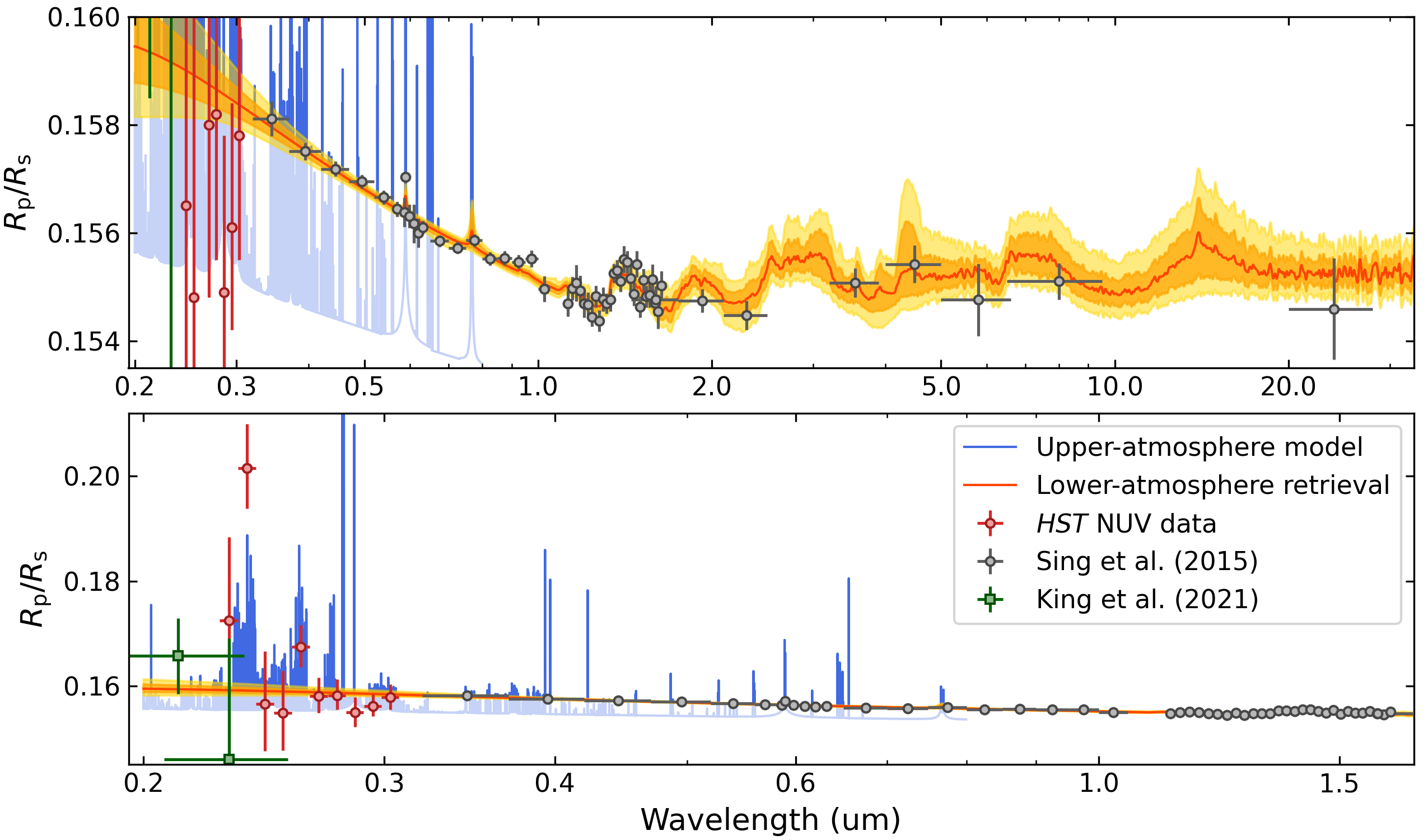}
\caption{Infrared-to-ultraviolet transmission spectrum of
  {\oneeightnineb}. Top and bottom panels show the same data, focused
  on different spectral regions for better visibility.  The colored
  markers with error bars show the transit observations, 1$\sigma$
  uncertainties, and wavelength span of the observations (see legend).
  Our data points have been computed from the data binned at a
  resolving power of R=33 (see Appendix
  \ref{sec:coarse_lightcurves}). The red curve and orange shaded areas
  shows the spectrum posterior distribution of the optical-IR fit
  (median, 68\%, and 95\% quantiles of the distribution, see Appendix
  \ref{sec:retrieval}).  The blue curve shows the upper-atmosphere
  model (shaded in a slightly clear color when sitting below the
  optical/infrared continuum).}
\label{fig:context}
\end{figure*}

Figures \ref{fig:coarse_transmission} and \ref{fig:context} show the
transmission spectrum from the upper atmosphere model in blue. Here,
we calibrated the model spectrum so that the measured base {\rprs} (in
the near-IR) corresponds to 2.3 mbar in the atmosphere
\citep{LavvasKoskinen2017apjAerosolProperties}. So as not to bias the
fit to any particular interpretation at this point, we did not include
extinction by the high-altitude haze modeled by
\citet{LavvasKoskinen2017apjAerosolProperties} in this forward
model. Instead, the transit continuum is due to Rayleigh scattering by
H$_2$, H, and He. It is immediately clear that the model continuum
falls below the extrapolation based on the optical-IR
transmission spectrum of the planet, due to the lack of the
high-altitude haze in the forward model. Since transit depth is not a
cumulative quantity, however, the model still correctly represents
absorption by Fe II and Mg II in the upper atmosphere in the absence
of condensation, even if high-altitude hazes are present.

We ran the upper atmosphere model assuming globally averaged radiative
forcing with the XUV spectrum constructed from $\epsilon$~Eridani as
an input, due to the well known similarity to HD~189733
\citep{LavvasKoskinen2017apjAerosolProperties}.  The reference
simulation, on which the spectrum in
Figure~\ref{fig:coarse_transmission} is based on, assumes an isolated
planet. We also ran a second simulation that includes Roche lobe
overflow \citep[see,][]{KoskinenEtal2022apjExtremePlanetsLoss} but this had
negligible effect on the modeled spectrum, in line with the
expectation that Roche lobe overflow does not significantly enhance
the escape rate or upper atmosphere densities on {\oneeightnineb}.

We note that NUV observations have the potential to offer valuable new
insights to the properties of possible high-altitude hazes and/or
clouds. Unfortunately, S/N is an issue for {\oneeightnineb} that
precludes clear conclusions on the origin of the continuum in the
transit observations. The visible spectrum shows a strong slope
indicative of small particle hazes
\citep{SingEtal2016natHotJupiterTransmission,
  LavvasKoskinen2017apjAerosolProperties} that appears to be supported
to some degree by observations by the Stratospheric Observatory for
Infrared Astronomy (SOFIA)
\citep{AngerhausenEtal2015jatisSOFIAtransitHD189733b}; however, the
latter found a slope offsetted to lower depths by
$\Delta\rprs= \sim0.001$ (see Appendix
  \ref{sec:coarse_lightcurves}).  The slope can, however, be enhanced
by stellar activity (star spots)
\citep{McCulloughEtal2014apjHD189733bHST}. Also, a general circulation
model (GCM) that simulates the production and transport of hazes
predicts a shallower slope unless mixing by sub-grid scale eddies
exceeds mixing by global circulation
\citep{SteinrueckEtal2021mnras3Dphotochemistry}. The NUV observations
are statistically consistent with the model that includes no haze, and
generally the observed transmission falls below the extrapolation
based on the visible slope (Figure \ref{fig:context}), with the
exception of the 2350 \AA~region. Due to low S/N in the NUV, however,
the observations are also consistent with this extrapolation. Thus,
these NUV observations neither confirm or exclude the presence of
high-altitude hazes, especially if the visible slope is enhanced by
star spots.

\subsection{Lower-atmosphere Modeling}
\label{sec:lower_model}

To constrain the lower-atmosphere properties of {\oneeightnineb} we
used the open-source {\pyratbay}
framework\footnote{\href{https://pyratbay.readthedocs.io/}
  {https://pyratbay.readthedocs.io/}}
\citep[][]{CubillosBlecic2021mnrasPyratBay}.  We modeled the
atmosphere from 100 to $10^{-9}$~bar, adopting a parametric
temperature profile from \citet{MadhusudhanSeager2009apjRetrieval},
hydrostatic equilibrium, and a H/He-dominated composition considering
constant-with-altitude volume mixing ratios.

The {\pyratbay} radiative transfer computed the transmission
spectrum between 0.2 and 33.0 ${\microns}$, at a constant resolving
power of $R=\lambda/\Delta\lambda=15,000$.  The opacities included
line-by-line data from HITEMP for {\carbdiox}, CO, and {\methane}
\citep{RothmanEtal2010jqsrtHITEMP, LiEtal2015apjsCOlineList,
  HargreavesEtal2020apjsHitempCH4}, and from ExoMol for HCN, {\water},
{\ammonia}, and SiO \citep{HarrisEtal2006mnrasHCNlineList,
  HarrisEtal2008mnrasExomolHCN,
  PolyanskyEtal2018mnrasPOKAZATELexomolH2O,
  ColesEtal2019mnrasNH3coyuteExomol,
  Yurchenko2015jqsrtBYTe15exomolNH3,
  YurchenkoEtal2022mnrasSIOUVENIRexomolSiO}. We extracted the dominant
transitions from these line-lists using the {\repack} algorithm
\citep{Cubillos2017apjRepack}, and then computed tabulated opacities
by sampling over temperature and wavelength grid.
Additionally, the model included collision-induced opacities for
{\molhyd}--{\molhyd} pairs \citep{BorysowEtal2001jqsrtH2H2highT,
  Borysow2002jqsrtH2H2lowT} and {\molhyd}--He pairs
\citep{BorysowEtal1988apjH2HeRT, BorysowEtal1989apjH2HeRVRT,
  BorysowFrommhold1989apjH2HeOvertones}; Rayleigh-scattering opacity
for H, {\molhyd}, and He \citep{Kurucz1970saorsAtlas}; and a gray
cloud deck model.  Finally, the model considered a parametric haze
absorber model \citep{LecavelierEtal2008aaRayleighHD189733b} with
arbitrary slope and strength ($\alpha_{\rm haze}$ and
$\kappa_{\rm haze}$), which are fit in the retrieval.

The atmospheric retrieval employed the differential-evolution
Markov-chain Monte Carlo sampler \citep{terBraak2008SnookerDEMC},
implemented via the open-source code {\mcc}
\citep{CubillosEtal2017apjRednoise}, guided by the transit
observations from {\HST}'s STIS G430 and G750, WFC3 G102 and G141
spectrographs; and from {\Spitzer}'s IRAC 3.6, 4.5, 5.8, 8.0 {$\micron$}
and MIPS 24 {$\micron$} photometric bands
\citep[][]{PontEtal2013mnrasHD189733b,
  SingEtal2016natHotJupiterTransmission}.

Figure \ref{fig:context} shows the fit to the optical-IR transit
observations.  Appendix \ref{sec:retrieval} presents the retrieval
posterior-distribution results for all model parameters.
The optical-IR observations point to a prominent haze slope, steeper
than a Rayleigh slope ($\alpha_{\rm haze} \approx -9.3 \pm 2.3$).
This is a similar conclusion to that found by previous retrieval
studies on this planet
\citep[e.g.,][]{PinhasEtal2019mnrasHotJupiterSampleRetrieval,
  Barstow2020mnrasCloudyRetrievals}. From a physical perspective, the
steep haze slope can be created by a combination of particle growth
and efficient eddy (turbulent) mixing
\citep{OhnoKawashima2020apjSuperRayleigh} or by stellar
inhomogeneities \citep{RackhamEtal2018apjStellarHeterogeneityI},
effects that were captured by the parametric haze model (rather than
being explicitly accounted for). The optical absorption is also
stronger than that produced by hydrogen (as in the upper-atmosphere
model), leading to a higher continuum level at optical wavelengths, as
well when extrapolated to NUV wavelengths. The NUV observations are
consistent with the optical-IR continuum model, with the exception of
the few data points that extend above the continuum (mainly the 2350
{\AA} region).  Lastly, we can see that the
constraint on the temperature profile becomes less precise above the
$\micro$bar level since optical/IR data is mostly probing the lower layers of
the atmosphere.  At the same time, at this high altitudes the retrieved temperature
profile starts to differ from the higher temperatures predicted by
self-consistent models that account for high-energy flux that heats the
thermosphere of the planet \citep[][]{LavvasKoskinen2017apjAerosolProperties},
i.e., the models used as the base of our upper-atmosphere model.

\section{On the Nature of the NUV Absorption}
\label{sec:speculation}

\subsection{{\feii} Absorption}

It is interesting that the excess NUV absorption matches precisely the
locations of the two strongest {\feii} bands in the region, while
there does not seem to be any absorption corresponding to the {\fei}
band shortly blueward of the magnesium doublet.  Unfortunately, the
low S/N achieved when binning the data at higher
resolution does not allow us to identify absorption from individual
lines.

In fact, a detection of {\feii} would be very controversial in this
planet for multiple reasons.  In first place, iron is not expected to
survive the cold trap at the lower layers.  Even if we neglected the
iron condensation, as is the case for our upper-atmosphere model, we
cannot match the large amplitude of the observed features when we bin
the model at the resolution of the observation.  Another argument
against {\feii} is that given the relatively cool expected temperature
of the planet and the late spectral type of the star, one would also
expect to observe {\fei} if {\feii} were detected.

If the NUV absorption indeed corresponded to {\feii}, it would imply
that our hydrodynamic model severely underestimates the absorption
signature, and we would be in need of a mechanism to enhance the lines
such that they become detectable in broader bins.  This would require
a significant broadening of the iron lines and super-solar abundances,
even when assuming no condensation in the lower atmosphere.  However,
global circulation models (GCM) predict winds of the order of
$\pm8$~km\,s$^{-1}$ at the terminator
\citep[][]{SteinrueckEtal2021mnras3Dphotochemistry}.  These models
include regions of significant size with flows both away from and
towards the Earth due to the complexity of the circulation and
rotation effects.  These kinds of flows could effectively broaden
metal lines in spatially unresolved transit observations.

\subsection{Absorption from an Unknown Absorber or other Source}

The broad-band absorption suggests the presence of an extra absorber,
but the fact that the absorption matched the location of {\feii} bands
might be deceptive.  It is certainly possible that the absorption
comes from another unidentified species, possibly from either a haze
or from the electronic band system of a molecular species, given the
width of the absorption and the transition energies involved.

Following \citet{LothringerEtal2022naturWASP178bSilicateUV}, we
explored whether SiO, say vaporized from strongly mixed cloud
particles, could appear in the upper atmosphere.  However, the atmospheric
retrievals do not favor significant SiO absorption when it was
included in the model.  From a further heuristic exploration, we could
not find any configuration of temperatures and abundances producing
SiO bands at the observed strength nor location.  We finally tested
for other candidates available in the ExoMol database and that
extended into the NUV, which led us to discard both HS and OH.

One might consider that the absorption originates
from an external source such as accreted material from trojan or
circumplanetary satellites
\citep[][]{KislyakovaEtal2016mnrasTrojanInfall,
  OzaEtal2019apjVolcanicNaK}. However, this is a highly speculative
scenario, since to date there is no model that demonstrates that
material from an external source can produce such large absorption
features, nor that it could generate exclusively an {\feii} signature
but not that from other particles like magnesium.

It seems also unlikely that the absorption features come from a
yet-undetermined aerosol. To our knowledge there is no self-consistent
GCM that couples haze production and circulation to a degree that it
can state with confidence that particles could form and exist at the
required high altitudes ($\sim$0.1 nbar). It is hard then to explain
how aerosols would form and remain aloft at such altitudes.

We also tested whether the stellar spot correction could generate the
observed features \citep[][]{RackhamEtal2018apjStellarHeterogeneityI,
  RackhamEtal2019ajStellarHeterogeneityII}, but on close inspection of
the spectral dependency of the spot contamination at different
temperatures, we found no indication that it could mimic extra
absorption at the location of the {\feii} bands.

Finally, we considered the impact of stellar center-to-limb flux
variation.  As the stellar {\feii} lines show in emission (Fig.\
\ref{fig:stellar_spectrum}), the observations should be probing the
(hotter) stellar chromosphere and transition region rather than the
(cooler) photosphere.  Therefore the lines may be showing limb
brightening instead of limb darkening, leading to a quite different
transit signature \citep[see,
e.g.,][]{SchlawinEtal2010apjLimbBrightening}.
Unfortunately, the limb-darkening/brightening law for these lines is
unknown, and it cannot be reasonably estimated with current
instrumentation.
In any case, while the unknown limb-darkening law of the emission
lines introduces a degree of uncertainty in the NUV transit depths,
unaccounted limb brightening would lead to underestimates transit
depths, and thus does not explain the excess absorption seen at
2350~{\AA} and 2600~{\AA}.

\subsection{Exoplanet NUV Observations in Context}
\label{sec:comparison}

To date three exoplanets have been spectroscopically observed during
transit in the NUV with {\HST}/STIS: WASP-121b
\citep{SingEtal2019ajWASP121bTransmissionNUV}, {\twoohnineb}
\citep{CubillosEtal2020ajHD209458bNUV}, and
{\oneeightnineb} (this work).  These three planets span a wide range of irradiation
regimes, receiving a stellar flux of 364$\times$ ({\oneeightnineb}),
769$\times$ ({\twoohnineb}), and 5624$\times$ (WASP-121b) that of the one
received by the Earth.
The analysis of this aggregated data provides a first glance at the
physical processes that shape the properties of the planets' upper
atmospheres.

\begin{figure}[t]
\centering
\includegraphics[width=\linewidth]{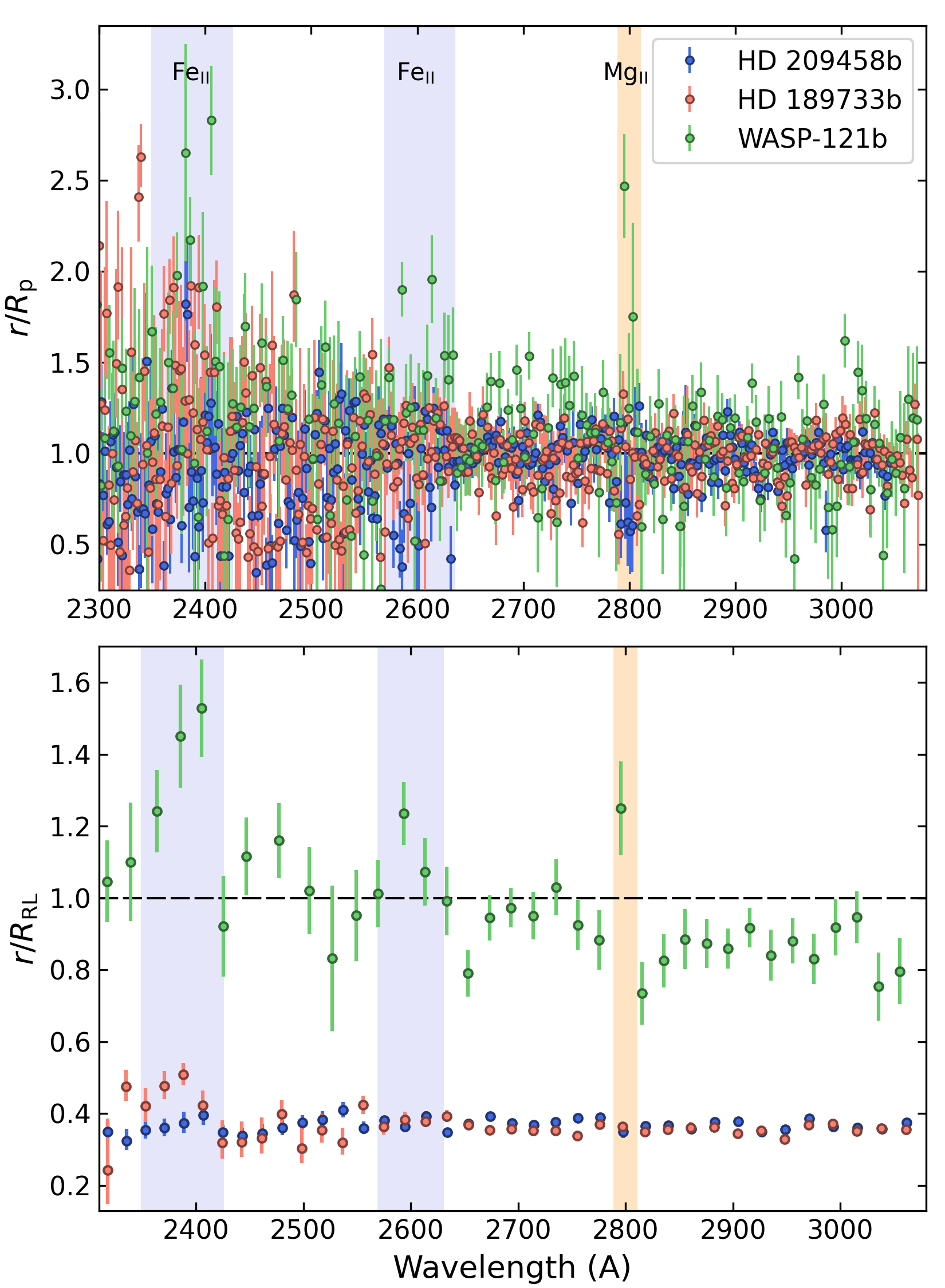}
\caption{Comparison {\HST}/STIS NUV transmission spectra
  of {\oneeightnineb},
  {\twoohnineb}, and WASP-121b scaled by the optical transit
  radius $\Rp$ (top panel) and by the Roche-lobe terminator
  radius $R_{\rm RL}$ of each planet (bottom panel).
  The shaded regions denote the wavelength span of the dominant
  {\feii} and {\mgii} NUV bands and lines.
  }
\label{fig:comparison}
\end{figure}

The upper panel of Fig.\ \ref{fig:comparison} shows the transmission
spectra
of these three planets at a resolution of $R=1300$ in units of
$r/R_{\rm p}$, with $R_{\rm p}$ being the optical radius.  We can see that
WASP-121b exhibits enhanced transit depths
coinciding with {\feii} and {\mgii} lines, whereas the other two planets
show subtler features.
When comparing the transmission spectra at a resolving power of $R=130$ in units of $r/R_{\rm RL}$,
with $R_{\rm RL}$ being the Roche-lobe terminator radius,
(lower panel of Fig.\ \ref{fig:comparison}) we see that the
pseudo-continuum extends to the Roche-lobe boundary, which indicates that
the middle atmosphere/lower thermosphere of WASP-121b undergoes Roche-lobe
overflow.  In clear contrast, this is not the case for
{\oneeightnineb} nor {\twoohnineb}.  WASP-121b is
therefore unique among these planets, and must have a much higher mass
loss rate than the others, as expected.  This broad behavior is
qualitatively consistent with expectations based on model predictions.

The lower panel of Fig. \ref{fig:comparison} illustrates the dramatic
difference between WASP-121b and the two other planets.  The NUV
spectra of {\twoohnineb} and  {\oneeightnineb} look nearly
featureless when compared to that of WASP-121b.  Certainly, WASP-121b
is not an ordinary planet---it must have substantially larger
mass-loss rates than the other two planets.  Neither {\twoohnineb} nor
{\oneeightnineb} show the {\mgii} absorption feature that survives in
the spectrum of WASP-121b.  As we
already know, the spectrum of {\oneeightnineb} shows a strong feature
close to the blue {\feii} absorption band. This feature does not extend to the Roche
lobe, but the peak is $\sim$150 surface scale heights above the
optical radius.  Given that heating in the thermosphere increases the
scale height by a factor of $\sim$20 above the $\sim$1 $\micro$bar
level, the peak probably probes the 1 nbar level.

Finally, we note that the feature in the spectrum of {\oneeightnineb}
is not identical to the {\feii} feature in the spectrum of WASP-121b
though.  In particular, the point at $\sim$2400 {\AA} coincides with a
strong {\feii} band and should show maximum absorption towards the longer-wavelength edge.  This
is true for WASP-121b, but in the spectrum of {\oneeightnineb}
absorption drops instead.  This poses a conundrum, however,
given the uncertainties in the data analysis, noise could be
responsible for this apparent inconsistency.

\section{Conclusions}
\label{sec:conclusions}

{\oneeightnineb} is the coolest hot Jupiter that has been
spectrosopically observed in the NUV so far. Transmission observations
indicate that this is a very hazy planet, but the composition of the
hazes are not really know, some of them may present additional
absorption in the NUV.

By analyzing the combined NUV {\HST}/STIS transmission spectra (3
transits) at a relatively coarse resolution ($R=50$), we obtained
better S/Ns and spectral resolution than previous NUV observation of
{\oneeightnineb} in the NUV by
\citet{KingEtal2021mnrasNUVxmmHD189733b}. The transit signature in the
light curves is clear and with no significant residual systematics.
We were able to constrain a continuum level that sits in between the
extrapolation of the optical haze slope and the continuum of a
pure-{\molhyd}/He upper atmosphere model.  Thus, given the
uncertainties of our continuum data points, the NUV data is
statistically consistent with either a hazy or clear atmosphere
scenario.
In addition, the NUV transmission spectrum
presents strong and broad absorption features that coincide with the
location of two strong {\feii} bands, while there is no apparent
excess absorption correlated with {\fei} bands.  When analyzing the
data at a higher resolution ($R=4700$), we found that most of the
lightcurves are dominated by noise, which does not allow us to confirm
nor discard iron as the source of the absorption.
However, thanks to the much higher stellar flux at the location of the
magnesium resonance lines, we were able to rule out the presense of
{\mgii} excess absorption above the continuum.  When compared to a
solar-metallicity model with magnesium in the upper atmosphere, the
measured transmission {\rprs} values at the core of the {\mgii} lines
are 2--4$\sigma$ below the model predictions.  The S/Ns of the data at
the {\mgi} line is not high enough to reject or confirm excess absorption
due to magnesium.

{\oneeightnineb} has a surface gravitational acceleration twice that
of {\twoohnineb} and unlike WASP-121b, Roche-lobe overflow effects are
not expected to be that significant. Its atmosphere is also cooler,
suggesting that the formation of clouds is expected. This is well in
agreement with optical observations, where the data are consistent
with a high-altitude haze, which could have a soot-like precursor.
{\oneeightnineb} upper-atmosphere models assuming that magnesium did
not condense in the lower atmosphere show considerable excess
absorption at the core of the {\mgii} resonance lines.  Thus, the
absense of excess absorption suggests that magnesium is not escaping
on {\oneeightnineb} and that Mg-Si clouds likely form in the lower
atmosphere.

At this point, the nature of the broad NUV absorption features is not
known.  An iron origin raises significant challenges for existing
upper-atmosphere models. In first place, given the relatively low
temperatures of {\oneeightnineb}, condensation is expected to
sequester heavy metals into the lower layers of the atmosphere,
though it is possible that condensates form primarily from other
metals instead, such as magnesium 
\citep{GaoEtal2020natasAerosolComposition,
  WoitkeEtal2020aaDiffusiveDustBrownDwarfs}.
However, even if iron is not strongly depleted, one would still need a
mechanism to enhance its absorption beyond that predicted by our
upper-atmosphere models. The two most immediate possibilities would be
to consider higher metallicities and zonal-wind velocity broadening of
the absorption lines. GCM estimations of {\oneeightnineb} predict wind
velocities at $\sim$8~km\,s$^{-1}$ at the terminator, the question for
future studies is whether these mechanisms are enough to boost the
signal.
Heavy metals play a key role in the modelling and interpretation of
NUV transit observations and hence in our understanding of
exoplanetary atmospheres as a whole.  The detection of metals at Roche
lobe distances would be surprising instead of expected for
{\oneeightnineb}.  Here we presented the detection of a broad
absorption feature in the NUV that matches the location of {\feii}
bands, but cannot fully confirm the nature of the absorber, both future
observational and theoretical work are needed to understand more
precisely the origin of this absorption.

\begin{acknowledgements}
We thank the anonymous referee for his/her time and valuable comments.
We thank contributors to the Python Programming
Language and the free and open-source community, including:
{\textsc{Pyrat Bay}} \citep{CubillosBlecic2021mnrasPyratBay},
{\mcc} \citep{CubillosEtal2017apjRednoise},
{\textsc{repack}} \citep{Cubillos2017apjRepack},
{\textsc{Numpy}} \citep{HarrisEtal2020natNumpy},
{\textsc{SciPy}} \citep{VirtanenEtal2020natmeScipy},
{\textsc{Matplotlib}} \citep{Hunter2007ieeeMatplotlib},
{\textsc{IPython}} \citep{PerezGranger2007cseIPython}, and
{\textsc{bibmanager}} \citep{Cubillos2020zndoBibmanager}.  This
research has made use of NASA's Astrophysics Data System Bibliographic
Services.
This project was funded in part by the Austrian Science Fund (FWF)
Erwin Schroedinger Fellowship, programs J4595-N and J4596-N.
A.G.S. and L.F. acknowledge financial support from the FFG project
865968.
Based on observations made with the NASA/ESA Hubble Space Telescope,
obtained at the Space Telescope Science Institute, which is operated
by the Association of Universities for Research in Astronomy, Inc.,
under NASA contract NAS 5-26555.  These observations are associated
with program \#15338.

\end{acknowledgements}

\bibliography{nuv_HD189733b}

\begin{thebibliography}{102}
\expandafter\ifx\csname natexlab\endcsname\relax\def\natexlab#1{#1}\fi
\ifx\adsurl \undefined \def\adsurl#1{\href{#1}{ADS}}\fi
\ifx\eprint \undefined \def\eprint#1{\href{http://arxiv.org/abs/#1}{#1}}\fi

\bibitem[{{Agol} {et~al.}(2010){Agol}, {Cowan}, {Knutson}, {Deming}, {Steffen},
  {Henry}, \& {Charbonneau}}]{AgolEtal2010apjHD189733b}
{Agol}, E. {et~al.} 2010, \apj, 721, 1861,
  \adsurl{https://ui.adsabs.harvard.edu/abs/2010ApJ...721.1861A},
  \eprint{1007.4378}

\bibitem[{{Andrae}(2010)}]{Andrae2010arxivErrorEstimation}
{Andrae}, R. 2010, arXiv e-prints, arXiv:1009.2755,
  \adsurl{https://ui.adsabs.harvard.edu/abs/2010arXiv1009.2755A},
  \eprint{1009.2755}

\bibitem[{{Angerhausen} {et~al.}(2015){Angerhausen}, {Mandushev}, {Mandell},
  {Dunham}, {Becklin}, {Collins}, {Hamilton}, {Logsdon}, {McElwain}, {McLean},
  {Pf{\"u}ller}, {Savage}, {Shenoy}, {Vacca}, {van Cleve}, \&
  {Wolf}}]{AngerhausenEtal2015jatisSOFIAtransitHD189733b}
{Angerhausen}, D. {et~al.} 2015, Journal of Astronomical Telescopes,
  Instruments, and Systems, 1, 034002,
  \adsurl{https://ui.adsabs.harvard.edu/abs/2015JATIS...1c4002A},
  \eprint{1507.01866}

\bibitem[{{Barnes} {et~al.}(2016){Barnes}, {Haswell}, {Staab}, \&
  {Anglada-Escud{\'e}}}]{BarnesEtal2016mnrasExcessAbsorptionHD189733b}
{Barnes}, J.~R., {Haswell}, C.~A., {Staab}, D., \& {Anglada-Escud{\'e}}, G.
  2016, \mnras, 462, 1012,
  \adsurl{https://ui.adsabs.harvard.edu/abs/2016MNRAS.462.1012B},
  \eprint{1607.03684}

\bibitem[{{Barstow}(2020)}]{Barstow2020mnrasCloudyRetrievals}
{Barstow}, J.~K. 2020, \mnras, 497, 4183,
  \adsurl{https://ui.adsabs.harvard.edu/abs/2020MNRAS.497.4183B},
  \eprint{2002.02945}

\bibitem[{{Ben-Jaffel} \&
  {Ballester}(2013)}]{BenJaffelBallester2013aaHD189733bOxygenHST}
{Ben-Jaffel}, L., \& {Ballester}, G.~E. 2013, \aap, 553, A52,
  \adsurl{https://ui.adsabs.harvard.edu/abs/2013A&A...553A..52B},
  \eprint{1303.4232}

\bibitem[{{Birkby} {et~al.}(2013){Birkby}, {de Kok}, {Brogi}, {de Mooij},
  {Schwarz}, {Albrecht}, \& {Snellen}}]{BirkbyEtal2013mnrasHD189733bHighRes}
{Birkby}, J.~L. {et~al.} 2013, \mnras, 436, L35,
  \adsurl{https://ui.adsabs.harvard.edu/abs/2013MNRAS.436L..35B},
  \eprint{1307.1133}

\bibitem[{{Bonomo} {et~al.}(2017){Bonomo}, {Desidera}, {Benatti}, {Borsa},
  {Crespi}, {Damasso}, {Lanza}, {Sozzetti}, {Lodato}, {Marzari}, {Boccato},
  {Claudi}, {Cosentino}, {Covino}, {Gratton}, {Maggio}, {Micela}, {Molinari},
  {Pagano}, {Piotto}, {Poretti}, {Smareglia}, {Affer}, {Biazzo}, {Bignamini},
  {Esposito}, {Giacobbe}, {H{\'e}brard}, {Malavolta}, {Maldonado}, {Mancini},
  {Martinez Fiorenzano}, {Masiero}, {Nascimbeni}, {Pedani}, {Rainer}, \&
  {Scandariato}}]{BonomoEtal2017aaRVmasses}
{Bonomo}, A.~S. {et~al.} 2017, \aap, 602, A107,
  \adsurl{https://ui.adsabs.harvard.edu/abs/2017A&A...602A.107B},
  \eprint{1704.00373}

\bibitem[{{Borysow}(2002)}]{Borysow2002jqsrtH2H2lowT}
{Borysow}, A. 2002, \aap, 390, 779,
  \adsurl{https://ui.adsabs.harvard.edu/abs/2002A&A...390..779B}

\bibitem[{{Borysow} \&
  {Frommhold}(1989)}]{BorysowFrommhold1989apjH2HeOvertones}
{Borysow}, A., \& {Frommhold}, L. 1989, \apj, 341, 549,
  \adsurl{https://ui.adsabs.harvard.edu/abs/1989ApJ...341..549B}

\bibitem[{{Borysow} {et~al.}(1989){Borysow}, {Frommhold}, \&
  {Moraldi}}]{BorysowEtal1989apjH2HeRVRT}
{Borysow}, A., {Frommhold}, L., \& {Moraldi}, M. 1989, \apj, 336, 495,
  \adsurl{https://ui.adsabs.harvard.edu/abs/1989ApJ...336..495B}

\bibitem[{{Borysow} {et~al.}(2001){Borysow}, {Jorgensen}, \&
  {Fu}}]{BorysowEtal2001jqsrtH2H2highT}
{Borysow}, A., {Jorgensen}, U.~G., \& {Fu}, Y. 2001, \jqsrt, 68, 235,
  \adsurl{https://ui.adsabs.harvard.edu/abs/2001JQSRT..68..235B}

\bibitem[{{Borysow} {et~al.}(1988){Borysow}, {Frommhold}, \&
  {Birnbaum}}]{BorysowEtal1988apjH2HeRT}
{Borysow}, J., {Frommhold}, L., \& {Birnbaum}, G. 1988, \apj, 326, 509,
  \adsurl{https://ui.adsabs.harvard.edu/abs/1988ApJ...326..509B}

\bibitem[{{Bouchy} {et~al.}(2005){Bouchy}, {Udry}, {Mayor}, {Moutou}, {Pont},
  {Iribarne}, {da Silva}, {Ilovaisky}, {Queloz}, {Santos}, {S{\'e}gransan}, \&
  {Zucker}}]{BouchyEtal2005aaHD189733discovery}
{Bouchy}, F. {et~al.} 2005, \aap, 444, L15,
  \adsurl{https://ui.adsabs.harvard.edu/abs/2005A&A...444L..15B},
  \eprint{astro-ph/0510119}

\bibitem[{{Bourrier} {et~al.}(2013){Bourrier}, {Lecavelier des Etangs},
  {Dupuy}, {Ehrenreich}, {Vidal-Madjar}, {H{\'e}brard}, {Ballester},
  {D{\'e}sert}, {Ferlet}, {Sing}, \&
  {Wheatley}}]{BourrierEtal2013aaHD189733bLyAlpha}
{Bourrier}, V. {et~al.} 2013, \aap, 551, A63,
  \adsurl{https://ui.adsabs.harvard.edu/abs/2013A&A...551A..63B},
  \eprint{1301.6030}

\bibitem[{{Brogi} {et~al.}(2016){Brogi}, {de Kok}, {Albrecht}, {Snellen},
  {Birkby}, \& {Schwarz}}]{BrogiEtal2016apjHD189733bHiresTransmission}
{Brogi}, M. {et~al.} 2016, \apj, 817, 106,
  \adsurl{https://ui.adsabs.harvard.edu/abs/2016ApJ...817..106B},
  \eprint{1512.05175}

\bibitem[{{Brogi} {et~al.}(2018){Brogi}, {Giacobbe}, {Guilluy}, {de Kok},
  {Sozzetti}, {Mancini}, \& {Bonomo}}]{BrogiEtal2018aaHD189733bGianoH2O}
{Brogi}, M. {et~al.} 2018, \aap, 615, A16,
  \adsurl{https://ui.adsabs.harvard.edu/abs/2018A&A...615A..16B},
  \eprint{1801.09569}

\bibitem[{{Brown} {et~al.}(2001){Brown}, {Charbonneau}, {Gilliland}, {Noyes},
  \& {Burrows}}]{BrownEtal2001apjHD209458bHSTstis}
{Brown}, T.~M., {Charbonneau}, D., {Gilliland}, R.~L., {Noyes}, R.~W., \&
  {Burrows}, A. 2001, \apj, 552, 699,
  \adsurl{https://ui.adsabs.harvard.edu/abs/2001ApJ...552..699B},
  \eprint{astro-ph/0101336}

\bibitem[{{Cauley} {et~al.}(2015){Cauley}, {Redfield}, {Jensen}, {Barman},
  {Endl}, \& {Cochran}}]{CauleyEtal2015apjHD189733bHiResHydrogenBow}
{Cauley}, P.~W. {et~al.} 2015, \apj, 810, 13,
  \adsurl{https://ui.adsabs.harvard.edu/abs/2015ApJ...810...13C},
  \eprint{1507.05916}

\bibitem[{{Cauley} {et~al.}(2018){Cauley}, {Shkolnik}, {Llama}, {Bourrier}, \&
  {Moutou}}]{CauleyEtal2018ajStarPlanetInteractionHD189733b}
{Cauley}, P.~W., {Shkolnik}, E.~L., {Llama}, J., {Bourrier}, V., \& {Moutou},
  C. 2018, \aj, 156, 262,
  \adsurl{https://ui.adsabs.harvard.edu/abs/2018AJ....156..262C},
  \eprint{1810.05253}

\bibitem[{{Coles} {et~al.}(2019){Coles}, {Yurchenko}, \&
  {Tennyson}}]{ColesEtal2019mnrasNH3coyuteExomol}
{Coles}, P.~A., {Yurchenko}, S.~N., \& {Tennyson}, J. 2019, \mnras, 490, 4638,
  \adsurl{https://ui.adsabs.harvard.edu/abs/2019MNRAS.490.4638C},
  \eprint{1911.10369}

\bibitem[{{Crouzet} {et~al.}(2014){Crouzet}, {McCullough}, {Deming}, \&
  {Madhusudhan}}]{CrouzetEtal2014apjHD189733bHSTwfc3EmissionH2O}
{Crouzet}, N., {McCullough}, P.~R., {Deming}, D., \& {Madhusudhan}, N. 2014,
  \apj, 795, 166,
  \adsurl{https://ui.adsabs.harvard.edu/abs/2014ApJ...795..166C},
  \eprint{1409.4000}

\bibitem[{{Cubillos} {et~al.}(2017){Cubillos}, {Harrington}, {Loredo}, {Lust},
  {Blecic}, \& {Stemm}}]{CubillosEtal2017apjRednoise}
{Cubillos}, P. {et~al.} 2017, \aj, 153, 3,
  \adsurl{https://ui.adsabs.harvard.edu/abs/2017AJ....153....3C},
  \eprint{1610.01336}

\bibitem[{{Cubillos}(2017)}]{Cubillos2017apjRepack}
{Cubillos}, P.~E. 2017, \apj, 850, 32,
  \adsurl{https://ui.adsabs.harvard.edu/abs/2017ApJ...850...32C},
  \eprint{1710.02556}

\bibitem[{{Cubillos}(2020)}]{Cubillos2020zndoBibmanager}
{Cubillos}, P.~E. 2020, {bibmanager: A BibTeX manager for LaTeX projects,
  Zenodo,
  doi:\href{https://zenodo.org/record/2547042}{10.5281/zenodo.2547042}},
  Zenodo,  \adsurl{https://ui.adsabs.harvard.edu/abs/2020zndo...2547042C}

\bibitem[{{Cubillos} \& {Blecic}(2021)}]{CubillosBlecic2021mnrasPyratBay}
{Cubillos}, P.~E., \& {Blecic}, J. 2021, \mnras, 505, 2675,
  \adsurl{https://ui.adsabs.harvard.edu/abs/2021MNRAS.505.2675C},
  \eprint{2105.05598}

\bibitem[{{Cubillos} {et~al.}(2020){Cubillos}, {Fossati}, {Koskinen}, {Young},
  {Salz}, {France}, {Sreejith}, \& {Haswell}}]{CubillosEtal2020ajHD209458bNUV}
{Cubillos}, P.~E. {et~al.} 2020, \aj, 159, 111,
  \adsurl{https://ui.adsabs.harvard.edu/abs/2020AJ....159..111C},
  \eprint{2001.03126}

\bibitem[{{D{\'e}sert} {et~al.}(2009){D{\'e}sert}, {Lecavelier des Etangs},
  {H{\'e}brard}, {Sing}, {Ehrenreich}, {Ferlet}, \&
  {Vidal-Madjar}}]{DesertEtal2009apjHD189733b}
{D{\'e}sert}, J.-M. {et~al.} 2009, \apj, 699, 478,
  \adsurl{https://ui.adsabs.harvard.edu/abs/2009ApJ...699..478D},
  \eprint{0903.3405}

\bibitem[{{Espinoza} \&
  {Jord{\'a}n}(2015)}]{EspinozaJordan2015mnrasLimbDarkeningI}
{Espinoza}, N., \& {Jord{\'a}n}, A. 2015, \mnras, 450, 1879,
  \adsurl{https://ui.adsabs.harvard.edu/abs/2015MNRAS.450.1879E},
  \eprint{1503.07020}

\bibitem[{{Gao} {et~al.}(2020){Gao}, {Thorngren}, {Lee}, {Fortney}, {Morley},
  {Wakeford}, {Powell}, {Stevenson}, \&
  {Zhang}}]{GaoEtal2020natasAerosolComposition}
{Gao}, P. {et~al.} 2020, Nature Astronomy, 4, 951,
  \adsurl{https://ui.adsabs.harvard.edu/abs/2020NatAs...4..951G},
  \eprint{2005.11939}

\bibitem[{{Guilluy} {et~al.}(2020){Guilluy}, {Andretta}, {Borsa}, {Giacobbe},
  {Sozzetti}, {Covino}, {Bourrier}, {Fossati}, {Bonomo}, {Esposito},
  {Giampapa}, {Harutyunyan}, {Rainer}, {Brogi}, {Bruno}, {Claudi}, {Frustagli},
  {Lanza}, {Mancini}, {Pino}, {Poretti}, {Scandariato}, {Affer}, {Baffa},
  {Baruffolo}, {Benatti}, {Biazzo}, {Bignamini}, {Boschin}, {Carleo},
  {Cecconi}, {Cosentino}, {Damasso}, {Desidera}, {Falcini}, {Martinez
  Fiorenzano}, {Ghedina}, {Gonz{\'a}lez-{\'A}lvarez}, {Guerra}, {Hernandez},
  {Leto}, {Maggio}, {Malavolta}, {Maldonado}, {Micela}, {Molinari},
  {Nascimbeni}, {Pagano}, {Pedani}, {Piotto}, \&
  {Reiners}}]{GuilluyEtal2020aaGianoHD189733bHelium}
{Guilluy}, G. {et~al.} 2020, \aap, 639, A49,
  \adsurl{https://ui.adsabs.harvard.edu/abs/2020A&A...639A..49G},
  \eprint{2005.05676}

\bibitem[{{Guo} \& {Ben-Jaffel}(2016)}]{GuoBenJaffel2016apjEUVinfluence}
{Guo}, J.~H., \& {Ben-Jaffel}, L. 2016, \apj, 818, 107,
  \adsurl{https://ui.adsabs.harvard.edu/abs/2016ApJ...818..107G},
  \eprint{1512.06470}

\bibitem[{{Hargreaves} {et~al.}(2020){Hargreaves}, {Gordon}, {Rey}, {Nikitin},
  {Tyuterev}, {Kochanov}, \& {Rothman}}]{HargreavesEtal2020apjsHitempCH4}
{Hargreaves}, R.~J. {et~al.} 2020, \apjs, 247, 55,
  \adsurl{https://ui.adsabs.harvard.edu/abs/2020ApJS..247...55H},
  \eprint{2001.05037}

\bibitem[{{Harris} {et~al.}(2020){Harris}, {Millman}, {van der Walt},
  {Gommers}, {Virtanen}, {Cournapeau}, {Wieser}, {Taylor}, {Berg}, {Smith},
  {Kern}, {Picus}, {Hoyer}, {van Kerkwijk}, {Brett}, {Haldane}, {del R{\'\i}o},
  {Wiebe}, {Peterson}, {G{\'e}rard-Marchant}, {Sheppard}, {Reddy}, {Weckesser},
  {Abbasi}, {Gohlke}, \& {Oliphant}}]{HarrisEtal2020natNumpy}
{Harris}, C.~R. {et~al.} 2020, \nat, 585, 357,
  \adsurl{https://ui.adsabs.harvard.edu/abs/2020Natur.585..357H},
  \eprint{2006.10256}

\bibitem[{{Harris} {et~al.}(2008){Harris}, {Larner}, {Tennyson}, {Kaminsky},
  {Pavlenko}, \& {Jones}}]{HarrisEtal2008mnrasExomolHCN}
{Harris}, G.~J. {et~al.} 2008, \mnras, 390, 143,
  \adsurl{https://ui.adsabs.harvard.edu/abs/2008MNRAS.390..143H},
  \eprint{0807.0717}

\bibitem[{{Harris} {et~al.}(2006){Harris}, {Tennyson}, {Kaminsky}, {Pavlenko},
  \& {Jones}}]{HarrisEtal2006mnrasHCNlineList}
{Harris}, G.~J., {Tennyson}, J., {Kaminsky}, B.~M., {Pavlenko}, Y.~V., \&
  {Jones}, H.~R.~A. 2006, \mnras, 367, 400,
  \adsurl{https://ui.adsabs.harvard.edu/abs/2006MNRAS.367..400H},
  \eprint{astro-ph/0512363}

\bibitem[{{Huang} {et~al.}(2017){Huang}, {Arras}, {Christie}, \&
  {Li}}]{HuangEtal2015apjHD189733bUpperModeling}
{Huang}, C., {Arras}, P., {Christie}, D., \& {Li}, Z.-Y. 2017, \apj, 851, 150,
  \adsurl{https://ui.adsabs.harvard.edu/abs/2017ApJ...851..150H},
  \eprint{1711.05428}

\bibitem[{{Huitson} {et~al.}(2012){Huitson}, {Sing}, {Vidal-Madjar},
  {Ballester}, {Lecavelier des Etangs}, {D{\'e}sert}, \&
  {Pont}}]{HuitsonEtal2012mnrasHD189bHSTstis}
{Huitson}, C.~M. {et~al.} 2012, \mnras, 422, 2477,
  \adsurl{https://ui.adsabs.harvard.edu/abs/2012MNRAS.422.2477H},
  \eprint{1202.4721}

\bibitem[{{Hunter}(2007)}]{Hunter2007ieeeMatplotlib}
{Hunter}, J.~D. 2007, Computing In Science \& Engineering, 9, 90

\bibitem[{{Husser} {et~al.}(2013){Husser}, {Wende-von Berg}, {Dreizler},
  {Homeier}, {Reiners}, {Barman}, \&
  {Hauschildt}}]{HusserEtal2013aaPHOENIXstellarModels}
{Husser}, T.~O. {et~al.} 2013, \aap, 553, A6,
  \adsurl{https://ui.adsabs.harvard.edu/abs/2013A&A...553A...6H},
  \eprint{1303.5632}

\bibitem[{{Jensen} {et~al.}(2012){Jensen}, {Redfield}, {Endl}, {Cochran},
  {Koesterke}, \& {Barman}}]{JensenEtalapj2012HalphaHD189733b}
{Jensen}, A.~G. {et~al.} 2012, \apj, 751, 86,
  \adsurl{https://ui.adsabs.harvard.edu/abs/2012ApJ...751...86J},
  \eprint{1203.4484}

\bibitem[{{King} {et~al.}(2021){King}, {Corrales}, {Wheatley}, {Lavvas},
  {Steinrueck}, {Bourrier}, {Ehrenreich}, {Lecavelier des Etangs}, \&
  {Louden}}]{KingEtal2021mnrasNUVxmmHD189733b}
{King}, G.~W. {et~al.} 2021, \mnras, 506, 2453,
  \adsurl{https://ui.adsabs.harvard.edu/abs/2021MNRAS.506.2453K},
  \eprint{2106.16208}

\bibitem[{{Kislyakova} {et~al.}(2016){Kislyakova}, {Pilat-Lohinger}, {Funk},
  {Lammer}, {Fossati}, {Eggl}, {Schwarz}, {Boudjada}, \&
  {Erkaev}}]{KislyakovaEtal2016mnrasTrojanInfall}
{Kislyakova}, K.~G. {et~al.} 2016, \mnras, 461, 988,
  \adsurl{https://ui.adsabs.harvard.edu/abs/2016MNRAS.461..988K},
  \eprint{1605.02507}

\bibitem[{{Knutson} {et~al.}(2007){Knutson}, {Charbonneau}, {Allen}, {Fortney},
  {Agol}, {Cowan}, {Showman}, {Cooper}, \&
  {Megeath}}]{KnutsonEtal2007nat8umPhaseHD189733b}
{Knutson}, H.~A. {et~al.} 2007, \nat, 447, 183,
  \adsurl{https://ui.adsabs.harvard.edu/abs/2007Natur.447..183K},
  \eprint{0705.0993}

\bibitem[{{Knutson} {et~al.}(2009){Knutson}, {Charbonneau}, {Cowan}, {Fortney},
  {Showman}, {Agol}, {Henry}, {Everett}, \&
  {Allen}}]{KnutsonEtal2009apj24umPhaseHD189733b}
{Knutson}, H.~A. {et~al.} 2009, \apj, 690, 822,
  \adsurl{https://ui.adsabs.harvard.edu/abs/2009ApJ...690..822K},
  \eprint{0802.1705}

\bibitem[{{Knutson} {et~al.}(2012){Knutson}, {Lewis}, {Fortney}, {Burrows},
  {Showman}, {Cowan}, {Agol}, {Aigrain}, {Charbonneau}, {Deming}, {D{\'e}sert},
  {Henry}, {Langton}, \&
  {Laughlin}}]{KnutsonEtal2012apj3.6um4.5umPhaseHD189733b}
{Knutson}, H.~A. {et~al.} 2012, \apj, 754, 22,
  \adsurl{https://ui.adsabs.harvard.edu/abs/2012ApJ...754...22K},
  \eprint{1206.6887}

\bibitem[{{Koskinen} {et~al.}(2013{\natexlab{a}}){Koskinen}, {Harris}, {Yelle},
  \& {Lavvas}}]{KoskinenEtal2013icarHD209458bMetalsEscapeI}
{Koskinen}, T.~T., {Harris}, M.~J., {Yelle}, R.~V., \& {Lavvas}, P.
  2013{\natexlab{a}}, \icarus, 226, 1678,
  \adsurl{https://ui.adsabs.harvard.edu/abs/2013Icar..226.1678K},
  \eprint{1210.1536}

\bibitem[{{Koskinen} {et~al.}(2022){Koskinen}, {Lavvas}, {Huang}, {Bergsten},
  {Fernandes}, \& {Young}}]{KoskinenEtal2022apjExtremePlanetsLoss}
{Koskinen}, T.~T. {et~al.} 2022, \apj, 929, 52,
  \adsurl{https://ui.adsabs.harvard.edu/abs/2022ApJ...929...52K},
  \eprint{2203.06302}

\bibitem[{{Koskinen} {et~al.}(2013{\natexlab{b}}){Koskinen}, {Yelle}, {Harris},
  \& {Lavvas}}]{KoskinenEtal2013icarHD209458bMetalsEscapeII}
{Koskinen}, T.~T., {Yelle}, R.~V., {Harris}, M.~J., \& {Lavvas}, P.
  2013{\natexlab{b}}, \icarus, 226, 1695,
  \adsurl{https://ui.adsabs.harvard.edu/abs/2013Icar..226.1695K},
  \eprint{1210.1543}

\bibitem[{{Kreidberg} {et~al.}(2014){Kreidberg}, {Bean}, {D{\'e}sert},
  {Benneke}, {Deming}, {Stevenson}, {Seager}, {Berta-Thompson}, {Seifahrt}, \&
  {Homeier}}]{KreidbergEtal2014natCloudsGJ1214b}
{Kreidberg}, L. {et~al.} 2014, \nat, 505, 69,
  \adsurl{https://ui.adsabs.harvard.edu/abs/2014Natur.505...69K},
  \eprint{1401.0022}

\bibitem[{{Kurucz}(1970)}]{Kurucz1970saorsAtlas}
{Kurucz}, R.~L. 1970, SAO Special Report, 309,
  \adsurl{https://ui.adsabs.harvard.edu/abs/1970SAOSR.309.....K}

\bibitem[{{Lavvas} \&
  {Koskinen}(2017)}]{LavvasKoskinen2017apjAerosolProperties}
{Lavvas}, P., \& {Koskinen}, T. 2017, \apj, 847, 32,
  \adsurl{https://ui.adsabs.harvard.edu/abs/2017ApJ...847...32L}

\bibitem[{{Lecavelier des Etangs} {et~al.}(2012){Lecavelier des Etangs},
  {Bourrier}, {Wheatley}, {Dupuy}, {Ehrenreich}, {Vidal-Madjar}, {H{\'e}brard},
  {Ballester}, {D{\'e}sert}, {Ferlet}, \&
  {Sing}}]{LecavelierEtal2012aaHD189733bEvapVariation}
{Lecavelier des Etangs}, A. {et~al.} 2012, \aap, 543, L4,
  \adsurl{https://ui.adsabs.harvard.edu/abs/2012A&A...543L...4L},
  \eprint{1206.6274}

\bibitem[{{Lecavelier Des Etangs} {et~al.}(2008){Lecavelier Des Etangs},
  {Pont}, {Vidal-Madjar}, \& {Sing}}]{LecavelierEtal2008aaRayleighHD189733b}
{Lecavelier Des Etangs}, A., {Pont}, F., {Vidal-Madjar}, A., \& {Sing}, D.
  2008, \aap, 481, L83,
  \adsurl{https://ui.adsabs.harvard.edu/abs/2008A&A...481L..83L},
  \eprint{0802.3228}

\bibitem[{{Li} {et~al.}(2015){Li}, {Gordon}, {Rothman}, {Tan}, {Hu}, {Kassi},
  {Campargue}, \& {Medvedev}}]{LiEtal2015apjsCOlineList}
{Li}, G. {et~al.} 2015, \apjs, 216, 15,
  \adsurl{https://ui.adsabs.harvard.edu/abs/2015ApJS..216...15L}

\bibitem[{{Liddle}(2007)}]{Liddle2007mnrasBIC}
{Liddle}, A.~R. 2007, \mnras, 377, L74,
  \adsurl{https://ui.adsabs.harvard.edu/abs/2007MNRAS.377L..74L},
  \eprint{astro-ph/0701113}

\bibitem[{{Lothringer} {et~al.}(2022){Lothringer}, {Sing}, {Rustamkulov},
  {Wakeford}, {Stevenson}, {Nikolov}, {Lavvas}, {Spake}, \&
  {Winch}}]{LothringerEtal2022naturWASP178bSilicateUV}
{Lothringer}, J.~D. {et~al.} 2022, \nat, 604, 49,
  \adsurl{https://ui.adsabs.harvard.edu/abs/2022Natur.604...49L},
  \eprint{2204.03639}

\bibitem[{{Louden} \&
  {Wheatley}(2015)}]{LoudenWheatley2015apjHD189733bHiresSodiumWinds}
{Louden}, T., \& {Wheatley}, P.~J. 2015, \apjl, 814, L24,
  \adsurl{https://ui.adsabs.harvard.edu/abs/2015ApJ...814L..24L},
  \eprint{1511.03689}

\bibitem[{{Madhusudhan} \& {Seager}(2009)}]{MadhusudhanSeager2009apjRetrieval}
{Madhusudhan}, N., \& {Seager}, S. 2009, \apj, 707, 24,
  \adsurl{https://ui.adsabs.harvard.edu/abs/2009ApJ...707...24M},
  \eprint{0910.1347}

\bibitem[{{Mandel} \& {Agol}(2002)}]{MandelAgol2002apjLightcurves}
{Mandel}, K., \& {Agol}, E. 2002, \apjl, 580, L171,
  \adsurl{https://ui.adsabs.harvard.edu/abs/2002ApJ...580L.171M},
  \eprint{astro-ph/0210099}

\bibitem[{{McCullough} {et~al.}(2014){McCullough}, {Crouzet}, {Deming}, \&
  {Madhusudhan}}]{McCulloughEtal2014apjHD189733bHST}
{McCullough}, P.~R., {Crouzet}, N., {Deming}, D., \& {Madhusudhan}, N. 2014,
  \apj, 791, 55,
  \adsurl{https://ui.adsabs.harvard.edu/abs/2014ApJ...791...55M},
  \eprint{1407.2462}

\bibitem[{{Menou}(2012)}]{Menou2012apjMagneticScaling}
{Menou}, K. 2012, \apj, 745, 138,
  \adsurl{https://ui.adsabs.harvard.edu/abs/2012ApJ...745..138M},
  \eprint{1108.3592}

\bibitem[{{Moses} {et~al.}(2011){Moses}, {Visscher}, {Fortney}, {Showman},
  {Lewis}, {Griffith}, {Klippenstein}, {Shabram}, {Friedson}, {Marley}, \&
  {Freedman}}]{MosesEtal2011apjDissequilibriumHD209nHD189b}
{Moses}, J.~I. {et~al.} 2011, \apj, 737, 15,
  \adsurl{https://ui.adsabs.harvard.edu/abs/2011ApJ...737...15M},
  \eprint{1102.0063}

\bibitem[{{Ohno} \& {Kawashima}(2020)}]{OhnoKawashima2020apjSuperRayleigh}
{Ohno}, K., \& {Kawashima}, Y. 2020, \apjl, 895, L47,
  \adsurl{https://ui.adsabs.harvard.edu/abs/2020ApJ...895L..47O},
  \eprint{2005.08880}

\bibitem[{{Oza} {et~al.}(2019){Oza}, {Johnson}, {Lellouch}, {Schmidt},
  {Schneider}, {Huang}, {Gamborino}, {Gebek}, {Wyttenbach}, {Demory},
  {Mordasini}, {Saxena}, {Dubois}, {Moullet}, \&
  {Thomas}}]{OzaEtal2019apjVolcanicNaK}
{Oza}, A.~V. {et~al.} 2019, \apj, 885, 168,
  \adsurl{https://ui.adsabs.harvard.edu/abs/2019ApJ...885..168O},
  \eprint{1908.10732}

\bibitem[{{P\'erez} \& {Granger}(2007)}]{PerezGranger2007cseIPython}
{P\'erez}, F., \& {Granger}, B.~E. 2007, Computing in Science and Engineering,
  9, 21

\bibitem[{{Pillitteri} {et~al.}(2015){Pillitteri}, {Maggio}, {Micela},
  {Sciortino}, {Wolk}, \&
  {Matsakos}}]{PillitteriEtal2015apjHD189733fuvVariability}
{Pillitteri}, I. {et~al.} 2015, \apj, 805, 52,
  \adsurl{https://ui.adsabs.harvard.edu/abs/2015ApJ...805...52P},
  \eprint{1503.05590}

\bibitem[{{Pinhas} {et~al.}(2019){Pinhas}, {Madhusudhan}, {Gandhi}, \&
  {MacDonald}}]{PinhasEtal2019mnrasHotJupiterSampleRetrieval}
{Pinhas}, A., {Madhusudhan}, N., {Gandhi}, S., \& {MacDonald}, R. 2019, \mnras,
  482, 1485,  \adsurl{https://ui.adsabs.harvard.edu/abs/2019MNRAS.482.1485P},
  \eprint{1811.00011}

\bibitem[{{Polyansky} {et~al.}(2018){Polyansky}, {Kyuberis}, {Zobov},
  {Tennyson}, {Yurchenko}, \&
  {Lodi}}]{PolyanskyEtal2018mnrasPOKAZATELexomolH2O}
{Polyansky}, O.~L. {et~al.} 2018, \mnras, 480, 2597,
  \adsurl{https://ui.adsabs.harvard.edu/abs/2018MNRAS.480.2597P},
  \eprint{1807.04529}

\bibitem[{{Pont} {et~al.}(2008){Pont}, {Knutson}, {Gilliland}, {Moutou}, \&
  {Charbonneau}}]{PontEtal2008mnrasHD189733bHaze}
{Pont}, F., {Knutson}, H., {Gilliland}, R.~L., {Moutou}, C., \& {Charbonneau},
  D. 2008, \mnras, 385, 109,
  \adsurl{https://ui.adsabs.harvard.edu/abs/2008MNRAS.385..109P},
  \eprint{0712.1374}

\bibitem[{{Pont} {et~al.}(2013){Pont}, {Sing}, {Gibson}, {Aigrain}, {Henry}, \&
  {Husnoo}}]{PontEtal2013mnrasHD189733b}
{Pont}, F. {et~al.} 2013, \mnras, 432, 2917,
  \adsurl{https://ui.adsabs.harvard.edu/abs/2013MNRAS.432.2917P},
  \eprint{1210.4163}

\bibitem[{{Poppenhaeger} {et~al.}(2013){Poppenhaeger}, {Schmitt}, \&
  {Wolk}}]{PoppenhaegerEtal2013HD189733bChandra}
{Poppenhaeger}, K., {Schmitt}, J.~H.~M.~M., \& {Wolk}, S.~J. 2013, \apj, 773,
  62,  \adsurl{https://ui.adsabs.harvard.edu/abs/2013ApJ...773...62P},
  \eprint{1306.2311}

\bibitem[{{Rackham} {et~al.}(2018){Rackham}, {Apai}, \&
  {Giampapa}}]{RackhamEtal2018apjStellarHeterogeneityI}
{Rackham}, B.~V., {Apai}, D., \& {Giampapa}, M.~S. 2018, \apj, 853, 122,
  \adsurl{https://ui.adsabs.harvard.edu/abs/2018ApJ...853..122R},
  \eprint{1711.05691}

\bibitem[{{Rackham} {et~al.}(2019){Rackham}, {Apai}, \&
  {Giampapa}}]{RackhamEtal2019ajStellarHeterogeneityII}
{Rackham}, B.~V., {Apai}, D., \& {Giampapa}, M.~S. 2019, \aj, 157, 96,
  \adsurl{https://ui.adsabs.harvard.edu/abs/2019AJ....157...96R},
  \eprint{1812.06184}

\bibitem[{{Raftery}(1995)}]{Raftery1995BIC}
{Raftery}, A.~E. 1995, Sociological Methodology, 25, 111

\bibitem[{{Redfield} {et~al.}(2008){Redfield}, {Endl}, {Cochran}, \&
  {Koesterke}}]{RedfieldEtal2008apjHD189733b}
{Redfield}, S., {Endl}, M., {Cochran}, W.~D., \& {Koesterke}, L. 2008, \apjl,
  673, L87,  \adsurl{https://ui.adsabs.harvard.edu/abs/2008ApJ...673L..87R},
  \eprint{0712.0761}

\bibitem[{{Rodler} {et~al.}(2013){Rodler}, {K{\"u}rster}, \&
  {Barnes}}]{RodlerEtal2013mnrasHD189733bHiresCO}
{Rodler}, F., {K{\"u}rster}, M., \& {Barnes}, J.~R. 2013, \mnras, 432, 1980,
  \adsurl{https://ui.adsabs.harvard.edu/abs/2013MNRAS.432.1980R}

\bibitem[{{Rothman} {et~al.}(2010){Rothman}, {Gordon}, {Barber}, {Dothe},
  {Gamache}, {Goldman}, {Perevalov}, {Tashkun}, \&
  {Tennyson}}]{RothmanEtal2010jqsrtHITEMP}
{Rothman}, L.~S. {et~al.} 2010, \jqsrt, 111, 2139,
  \adsurl{https://ui.adsabs.harvard.edu/abs/2010JQSRT.111.2139R}

\bibitem[{{Salz} {et~al.}(2018){Salz}, {Czesla}, {Schneider}, {Nagel},
  {Schmitt}, {Nortmann}, {Alonso-Floriano}, {L{\'o}pez-Puertas}, {Lamp{\'o}n},
  {Bauer}, {Snellen}, {Pall{\'e}}, {Caballero}, {Yan}, {Chen}, {Sanz-Forcada},
  {Amado}, {Quirrenbach}, {Ribas}, {Reiners}, {B{\'e}jar}, {Casasayas-Barris},
  {Cort{\'e}s-Contreras}, {Dreizler}, {Guenther}, {Henning}, {Jeffers},
  {Kaminski}, {K{\"u}rster}, {Lafarga}, {Lara}, {Molaverdikhani}, {Montes},
  {Morales}, {S{\'a}nchez-L{\'o}pez}, {Seifert}, {Zapatero Osorio}, \&
  {Zechmeister}}]{SalzEtal2018aaHeHD189733b}
{Salz}, M. {et~al.} 2018, \aap, 620, A97,
  \adsurl{https://ui.adsabs.harvard.edu/abs/2018A&A...620A..97S},
  \eprint{1812.02453}

\bibitem[{{Schlawin} {et~al.}(2010){Schlawin}, {Agol}, {Walkowicz}, {Covey}, \&
  {Lloyd}}]{SchlawinEtal2010apjLimbBrightening}
{Schlawin}, E., {Agol}, E., {Walkowicz}, L.~M., {Covey}, K., \& {Lloyd}, J.~P.
  2010, \apjl, 722, L75,
  \adsurl{https://ui.adsabs.harvard.edu/abs/2010ApJ...722L..75S},
  \eprint{1008.1073}

\bibitem[{{Showman} {et~al.}(2013){Showman}, {Fortney}, {Lewis}, \&
  {Shabram}}]{ShowmanEtal2013apjHotJupiterCirculation}
{Showman}, A.~P., {Fortney}, J.~J., {Lewis}, N.~K., \& {Shabram}, M. 2013,
  \apj, 762, 24,
  \adsurl{https://ui.adsabs.harvard.edu/abs/2013ApJ...762...24S},
  \eprint{1207.5639}

\bibitem[{{Showman} {et~al.}(2009){Showman}, {Fortney}, {Lian}, {Marley},
  {Freedman}, {Knutson}, \& {Charbonneau}}]{ShowmanEtal2009apjRadGCM}
{Showman}, A.~P. {et~al.} 2009, \apj, 699, 564,
  \adsurl{https://ui.adsabs.harvard.edu/abs/2009ApJ...699..564S},
  \eprint{0809.2089}

\bibitem[{{Shulyak} {et~al.}(2004){Shulyak}, {Tsymbal}, {Ryabchikova},
  {St{\"u}tz}, \& {Weiss}}]{ShulyakEtal2004aaLBLstellarModels}
{Shulyak}, D., {Tsymbal}, V., {Ryabchikova}, T., {St{\"u}tz}, C., \& {Weiss},
  W.~W. 2004, \aap, 428, 993,
  \adsurl{https://ui.adsabs.harvard.edu/abs/2004A&A...428..993S}

\bibitem[{{Sing} {et~al.}(2016){Sing}, {Fortney}, {Nikolov}, {Wakeford},
  {Kataria}, {Evans}, {Aigrain}, {Ballester}, {Burrows}, {Deming},
  {D{\'e}sert}, {Gibson}, {Henry}, {Huitson}, {Knutson}, {Lecavelier Des
  Etangs}, {Pont}, {Showman}, {Vidal-Madjar}, {Williamson}, \&
  {Wilson}}]{SingEtal2016natHotJupiterTransmission}
{Sing}, D.~K. {et~al.} 2016, \nat, 529, 59,
  \adsurl{https://ui.adsabs.harvard.edu/abs/2016Natur.529...59S},
  \eprint{1512.04341}

\bibitem[{{Sing} {et~al.}(2019){Sing}, {Lavvas}, {Ballester}, {Lecavelier des
  Etangs}, {Marley}, {Nikolov}, {Ben-Jaffel}, {Bourrier}, {Buchhave}, {Deming},
  {Ehrenreich}, {Mikal-Evans}, {Kataria}, {Lewis}, {L{\'o}pez-Morales},
  {Garc{\'\i}a Mu{\~n}oz}, {Henry}, {Sanz-Forcada}, {Spake}, {Wakeford}, \&
  {PanCET Collaboration}}]{SingEtal2019ajWASP121bTransmissionNUV}
{Sing}, D.~K. {et~al.} 2019, \aj, 158, 91,
  \adsurl{https://ui.adsabs.harvard.edu/abs/2019AJ....158...91S},
  \eprint{1908.00619}

\bibitem[{{Sing} {et~al.}(2011){Sing}, {Pont}, {Aigrain}, {Charbonneau},
  {D{\'e}sert}, {Gibson}, {Gilliland}, {Hayek}, {Henry}, {Knutson}, {Lecavelier
  Des Etangs}, {Mazeh}, \& {Shporer}}]{SingEtal2011mnrasHD189733b_STIS}
{Sing}, D.~K. {et~al.} 2011, \mnras, 416, 1443,
  \adsurl{https://ui.adsabs.harvard.edu/abs/2011MNRAS.416.1443S},
  \eprint{1103.0026}

\bibitem[{{Spiegel} {et~al.}(2009){Spiegel}, {Silverio}, \&
  {Burrows}}]{SpiegelEtal2009apjTiO}
{Spiegel}, D.~S., {Silverio}, K., \& {Burrows}, A. 2009, \apj, 699, 1487,
  \adsurl{https://ui.adsabs.harvard.edu/abs/2009ApJ...699.1487S},
  \eprint{0902.3995}

\bibitem[{{Stassun} {et~al.}(2017){Stassun}, {Collins}, \&
  {Gaudi}}]{StassunEtal2017ajGaiaRadiiMasses}
{Stassun}, K.~G., {Collins}, K.~A., \& {Gaudi}, B.~S. 2017, \aj, 153, 136,
  \adsurl{https://ui.adsabs.harvard.edu/abs/2017AJ....153..136S},
  \eprint{1609.04389}

\bibitem[{{Steinrueck} {et~al.}(2021){Steinrueck}, {Showman}, {Lavvas},
  {Koskinen}, {Tan}, \& {Zhang}}]{SteinrueckEtal2021mnras3Dphotochemistry}
{Steinrueck}, M.~E. {et~al.} 2021, \mnras, 504, 2783,
  \adsurl{https://ui.adsabs.harvard.edu/abs/2021MNRAS.504.2783S},
  \eprint{2011.14022}

\bibitem[{{ter Braak} \& {Vrugt}(2008)}]{terBraak2008SnookerDEMC}
{ter Braak}, C. J.~F., \& {Vrugt}, J.~A. 2008, Statistics and Computing, 18,
  435

\bibitem[{{Todorov} {et~al.}(2014){Todorov}, {Deming}, {Burrows}, \&
  {Grillmair}}]{TodorovEtal2014apjHD189733bSpitzerEmissionIRS}
{Todorov}, K.~O., {Deming}, D., {Burrows}, A., \& {Grillmair}, C.~J. 2014,
  \apj, 796, 100,
  \adsurl{https://ui.adsabs.harvard.edu/abs/2014ApJ...796..100T},
  \eprint{1410.1400}

\bibitem[{{Trotta}(2007)}]{Trotta2007mnrasBayesianModelSelection}
{Trotta}, R. 2007, \mnras, 378, 72,
  \adsurl{https://ui.adsabs.harvard.edu/abs/2007MNRAS.378...72T},
  \eprint{astro-ph/0504022}

\bibitem[{{Vidal-Madjar} {et~al.}(2008){Vidal-Madjar}, {Lecavelier des Etangs},
  {D{\'e}sert}, {Ballester}, {Ferlet}, {H{\'e}brard}, \&
  {Mayor}}]{VidalMadjarEtal2008apjHD209458bEvaporation}
{Vidal-Madjar}, A. {et~al.} 2008, \apjl, 676, L57,
  \adsurl{https://ui.adsabs.harvard.edu/abs/2008ApJ...676L..57V},
  \eprint{0802.0587}

\bibitem[{{Virtanen} {et~al.}(2020){Virtanen}, {Gommers}, {Oliphant},
  {Haberland}, {Reddy}, {Cournapeau}, {Burovski}, {Peterson}, {Weckesser},
  {Bright}, {van der Walt}, {Brett}, {Wilson}, {Millman}, {Mayorov}, {Nelson},
  {Jones}, {Kern}, {Larson}, {Carey}, {Polat}, {Feng}, {Moore}, {VanderPlas},
  {Laxalde}, {Perktold}, {Cimrman}, {Henriksen}, {Quintero}, {Harris},
  {Archibald}, {Ribeiro}, {Pedregosa}, {van Mulbregt}, \& {SciPy 1. 0
  Contributors}}]{VirtanenEtal2020natmeScipy}
{Virtanen}, P. {et~al.} 2020, Nature Methods, 17, 261,
  \adsurl{https://ui.adsabs.harvard.edu/abs/2020NatMe..17..261V},
  \eprint{1907.10121}

\bibitem[{{Wakeford} \& {Sing}(2015)}]{WakefordSing2015aaTransitCondensates}
{Wakeford}, H.~R., \& {Sing}, D.~K. 2015, \aap, 573, A122,
  \adsurl{https://ui.adsabs.harvard.edu/abs/2015A&A...573A.122W},
  \eprint{1409.7594}

\bibitem[{{Wakeford} {et~al.}(2016){Wakeford}, {Sing}, {Evans}, {Deming}, \&
  {Mandell}}]{WakefordEtal2016apjHSTsystematics}
{Wakeford}, H.~R., {Sing}, D.~K., {Evans}, T., {Deming}, D., \& {Mandell}, A.
  2016, \apj, 819, 10,
  \adsurl{https://ui.adsabs.harvard.edu/abs/2016ApJ...819...10W},
  \eprint{1601.02587}

\bibitem[{{Woitke} {et~al.}(2020){Woitke}, {Helling}, \&
  {Gunn}}]{WoitkeEtal2020aaDiffusiveDustBrownDwarfs}
{Woitke}, P., {Helling}, C., \& {Gunn}, O. 2020, \aap, 634, A23,
  \adsurl{https://ui.adsabs.harvard.edu/abs/2020A&A...634A..23W},
  \eprint{1911.03777}

\bibitem[{{Wyttenbach} {et~al.}(2015){Wyttenbach}, {Ehrenreich}, {Lovis},
  {Udry}, \& {Pepe}}]{WyttenbachEtal2015aaHD189733bSodiumHARPS}
{Wyttenbach}, A., {Ehrenreich}, D., {Lovis}, C., {Udry}, S., \& {Pepe}, F.
  2015, \aap, 577, A62,
  \adsurl{https://ui.adsabs.harvard.edu/abs/2015A&A...577A..62W},
  \eprint{1503.05581}

\bibitem[{{Yurchenko}(2015)}]{Yurchenko2015jqsrtBYTe15exomolNH3}
{Yurchenko}, S.~N. 2015, \jqsrt, 152, 28,
  \adsurl{https://ui.adsabs.harvard.edu/abs/2015JQSRT.152...28Y},
  \eprint{1502.07975}

\bibitem[{{Yurchenko} {et~al.}(2022){Yurchenko}, {Tennyson}, {Syme}, {Adam},
  {Clark}, {Cooper}, {Dobney}, {Donnelly}, {Gorman}, {Lynas-Gray}, {Meltzer},
  {Owens}, {Qu}, {Semenov}, {Somogyi}, {Upadhyay}, {Wright}, \& {Zapata
  Trujillo}}]{YurchenkoEtal2022mnrasSIOUVENIRexomolSiO}
{Yurchenko}, S.~N. {et~al.} 2022, \mnras, 510, 903,
  \adsurl{https://ui.adsabs.harvard.edu/abs/2022MNRAS.510..903Y},
  \eprint{2111.04859}

\bibitem[{{Zellem} {et~al.}(2017){Zellem}, {Swain}, {Roudier}, {Shkolnik},
  {Creech-Eakman}, {Ciardi}, {Line}, {Iyer}, {Bryden}, {Llama}, \&
  {Fahy}}]{ZellemEtal2017apjTransitStellarActivity}
{Zellem}, R.~T. {et~al.} 2017, \apj, 844, 27,
  \adsurl{https://ui.adsabs.harvard.edu/abs/2017ApJ...844...27Z},
  \eprint{1705.04708}

\bibitem[{{Zhang} {et~al.}(2022){Zhang}, {Cauley}, {Knutson}, {France},
  {Kreidberg}, {Oklop{\v{c}}i{\'c}}, {Redfield}, \&
  {Shkolnik}}]{ZhangEtal2022ajVariableHeliumHD189733b}
{Zhang}, M. {et~al.} 2022, \aj, 164, 237,
  \adsurl{https://ui.adsabs.harvard.edu/abs/2022AJ....164..237Z},
  \eprint{2204.02985}

\end{thebibliography}

\begin{appendix}

\section{{\oneeightnineb} NUV Transmission Spectra}
\label{sec:coarse_lightcurves}

\begin{figure}[h]
\centering
\includegraphics[width=\linewidth]{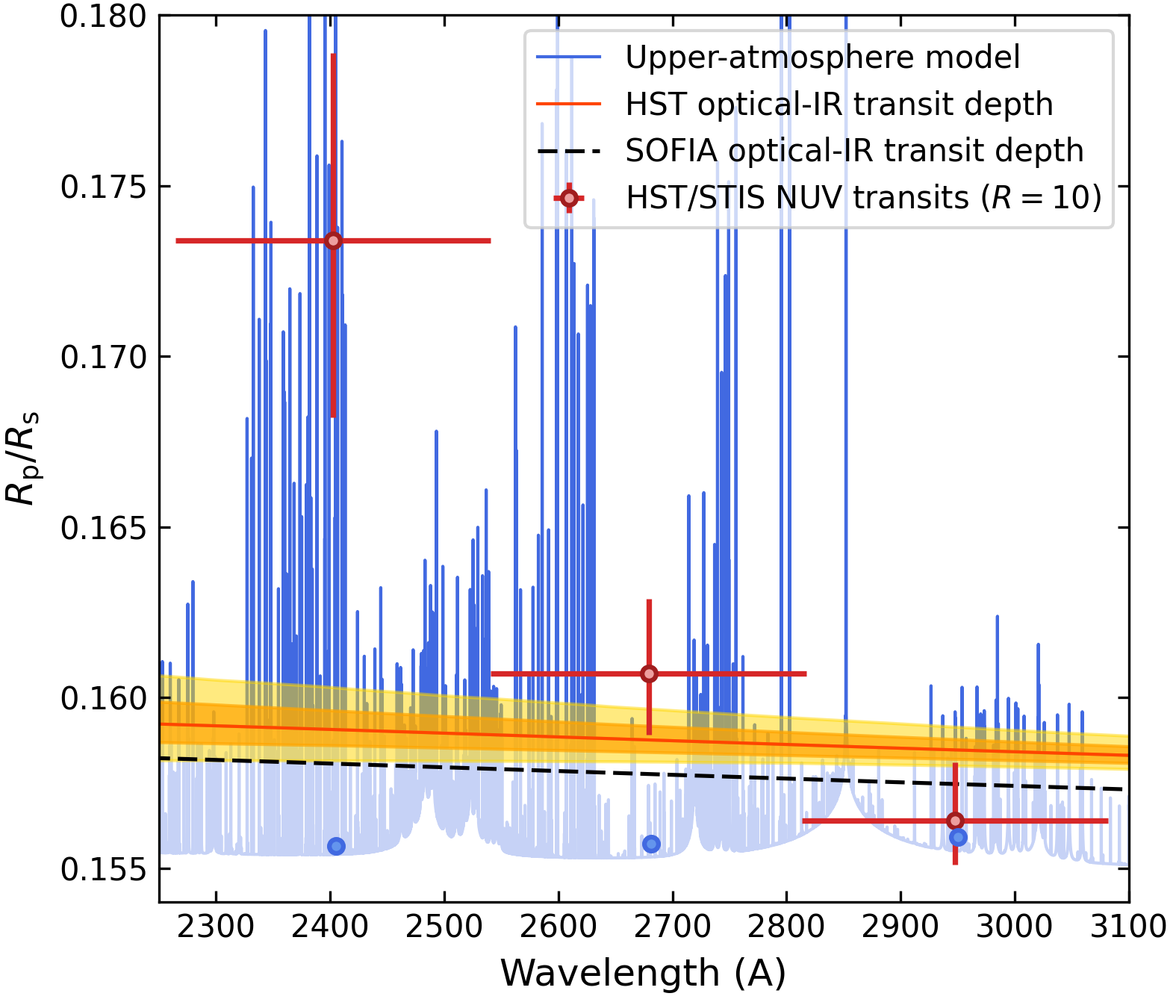}
\caption{NUV transmission spectrum of {\oneeightnineb} from the
  combined {\HST}/STIS observations.  The red markers with error bars
  denote the systematics- and stellar-spot corrected transmission
  spectrum, their 1$\sigma$ uncertainties, and the span of the
  spectral bins for a resolving power of R\,=\,10.  The blue spectrum
  shows a theoretical model of the planet's upper atmosphere.  The
  orange curve shows the fit to the optical-IR transmission spectrum
  extrapolated into the NUV (the shaded areas denote the 1$\sigma$ and
  2$\sigma$ uncertainties).  The black dashed curve shows the
  optical-IR fit shifted downwards according to a SOFIA-like offset of
  $\Delta \rprs = 0.001$.}
\label{fig:coarse_transmission_R10}
\end{figure}

Fig \ref{fig:coarse_transmission_R10} shows the NUV
  transmission spectrum binned at a coarse resolving power of $R = 10$
  and shows a SOFIA-like offset of the transit depth for comparison.

\begin{figure*}[h]
\centering
\includegraphics[width=\linewidth]{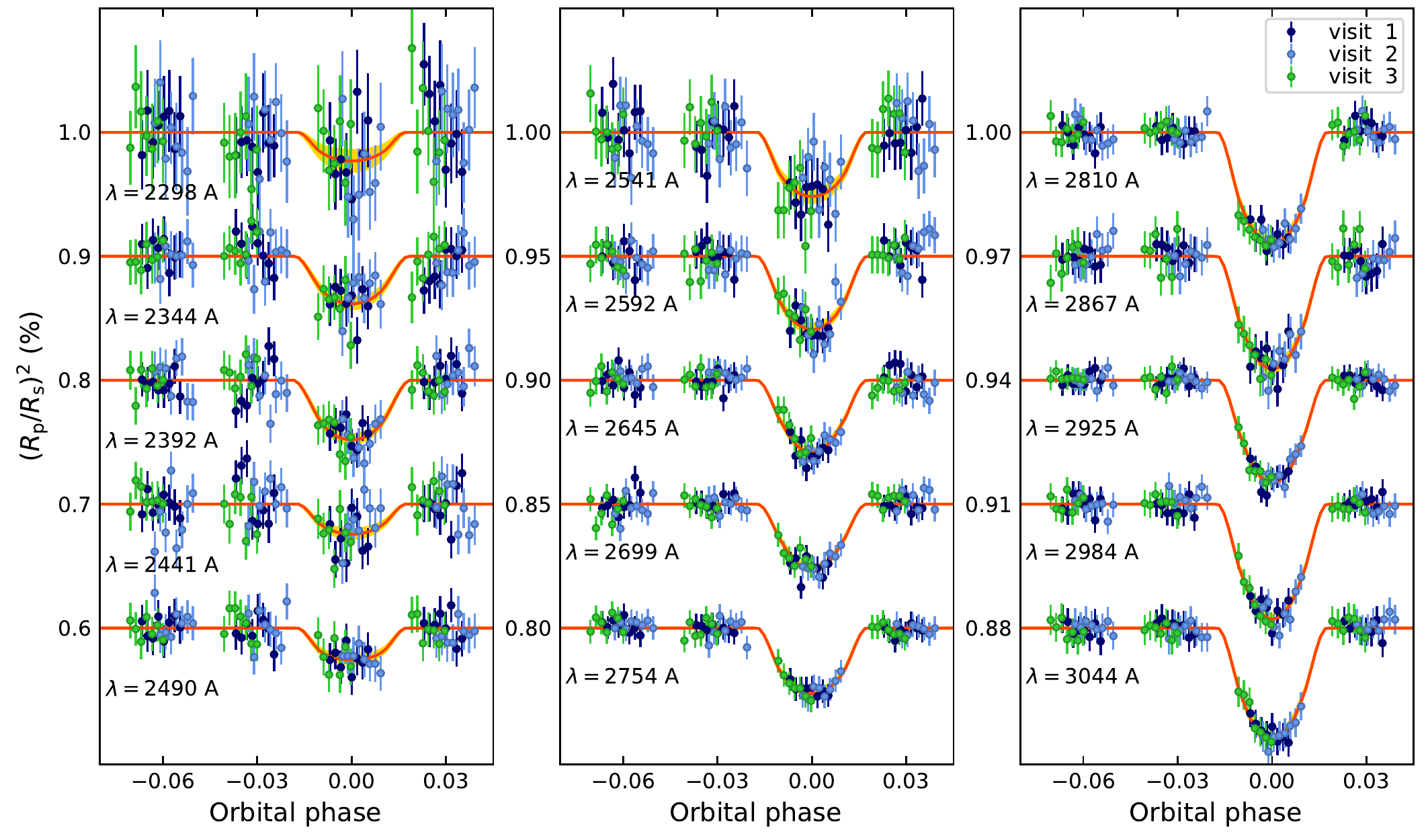}
\caption{Transit light curves of the combined {\HST}/STIS observations
  of {\oneeightnineb} binned at a resolving power of ${\rm R} = 50$.
  The colored markers with error bars denote the systematics- and
  stellar-spot corrected measurements, color coded for each visit (see
  legend).  The solid red curve and orange area denote the
  best-fitting transit depths and span of their 1$\sigma$
  uncertainties. The lightcurve of each bin has been shifted in the
  vertical axis for visualization.  The labels next to each ligtcurve
  denotes the mean wavelength of the spectral bin.  Note that the
  scale of the vertical axis changes from left to right as the S/N of
  the data improves with increasing wavelength.}
\label{fig:coarse_lightcurves}
\end{figure*}

Figure \ref{fig:coarse_lightcurves} shows the systematics- and
stellar-spot corrected lightcurves of the combined {\HST}/STIS
{\oneeightnineb} observations, when binned over a coarse wavelength
sampling with a resolving power of ${\rm R} = 50$.  We used the
posterior-distribution median and the central 68\% percentiles statistics
to estimate the parameter values and their credible intervals.

Tables \ref{table:coarse50}, \ref{table:coarse33},
  \ref{table:coarse10} present the estimated values of the NUV
  transmission planet-to-star radius ratio when analyzed at resolving
  powers of 50, 33, and 10, respectively.

{\renewcommand{\arraystretch}{1.5}
\begin{table}[h]
\centering
\caption{{\oneeightnineb} NUV Transmission Spectrum at R=50}
\label{table:coarse50}
\begin{tabular} {lccc}
\hline
\hline
$\lambda$ ($\microns$) & Bin half-width ($\microns$) & $\rprs$ \\
\hline
$2297.5$  & $ 23.1$  & $0.139_{-0.022}^{+0.036}$ \\
$2344.2$  & $ 23.6$  & $0.189_{-0.011}^{+0.013}$ \\
$2391.9$  & $ 24.1$  & $0.212_{-0.009}^{+0.009}$ \\
$2440.5$  & $ 24.6$  & $0.150_{-0.012}^{+0.014}$ \\
$2490.1$  & $ 25.1$  & $0.154_{-0.010}^{+0.011}$ \\
$2540.7$  & $ 25.6$  & $0.155_{-0.008}^{+0.009}$ \\
$2592.4$  & $ 26.1$  & $0.166_{-0.005}^{+0.006}$ \\
$2645.1$  & $ 26.6$  & $0.165_{-0.004}^{+0.004}$ \\
$2698.8$  & $ 27.2$  & $0.156_{-0.004}^{+0.004}$ \\
$2753.7$  & $ 27.7$  & $0.157_{-0.004}^{+0.004}$ \\
$2809.6$  & $ 28.3$  & $0.159_{-0.003}^{+0.003}$ \\
$2866.7$  & $ 28.8$  & $0.160_{-0.003}^{+0.004}$ \\
$2925.0$  & $ 29.4$  & $0.150_{-0.002}^{+0.003}$ \\
$2984.5$  & $ 30.0$  & $0.161_{-0.003}^{+0.003}$ \\
$3043.7$  & $ 29.2$  & $0.158_{-0.003}^{+0.003}$ \\
\hline
\end{tabular}
\end{table}
}

{\renewcommand{\arraystretch}{1.5}
\begin{table}[h]
\centering
\caption{{\oneeightnineb} NUV Transmission Spectrum at R=33}
\label{table:coarse33}
\begin{tabular} {lccc}
\hline
\hline
$\lambda$ ($\microns$) & Bin half-width ($\microns$) & $\rprs$ \\
\hline
$2309.4$  & $ 35.0$  & $0.172_{-0.014}^{+0.016}$ \\
$2380.5$  & $ 36.1$  & $0.202_{-0.008}^{+0.008}$ \\
$2453.7$  & $ 37.2$  & $0.157_{-0.009}^{+0.010}$ \\
$2529.2$  & $ 38.3$  & $0.155_{-0.007}^{+0.008}$ \\
$2607.1$  & $ 39.5$  & $0.168_{-0.004}^{+0.004}$ \\
$2687.3$  & $ 40.7$  & $0.158_{-0.003}^{+0.004}$ \\
$2770.0$  & $ 42.0$  & $0.158_{-0.003}^{+0.003}$ \\
$2855.2$  & $ 43.3$  & $0.155_{-0.003}^{+0.003}$ \\
$2943.0$  & $ 44.6$  & $0.156_{-0.002}^{+0.002}$ \\
$3030.3$  & $ 42.7$  & $0.158_{-0.002}^{+0.003}$ \\
\hline
\end{tabular}
\end{table}
}

{\renewcommand{\arraystretch}{1.5}
\begin{table}[h]
\centering
\caption{{\oneeightnineb} NUV Transmission Spectrum at R=10}
\label{table:coarse10}
\begin{tabular} {lccc}
\hline
\hline
$\lambda$ ($\microns$) & Bin half-width ($\microns$) & $\rprs$ \\
\hline
$2405.1$  & $130.7$ & $0.1734_{-0.0052}^{+0.0055}$ \\
$2681.6$  & $145.7$ & $0.1607_{-0.0018}^{+0.0022}$ \\
$2950.1$  & $122.8$ & $0.1564_{-0.0013}^{+0.0017}$ \\
\hline
\end{tabular}
\end{table}
}

\section{Lower-atmosphere Transmission Retrieval}
\label{sec:retrieval}

\begin{figure*}[h]
\centering
\includegraphics[width=\linewidth]{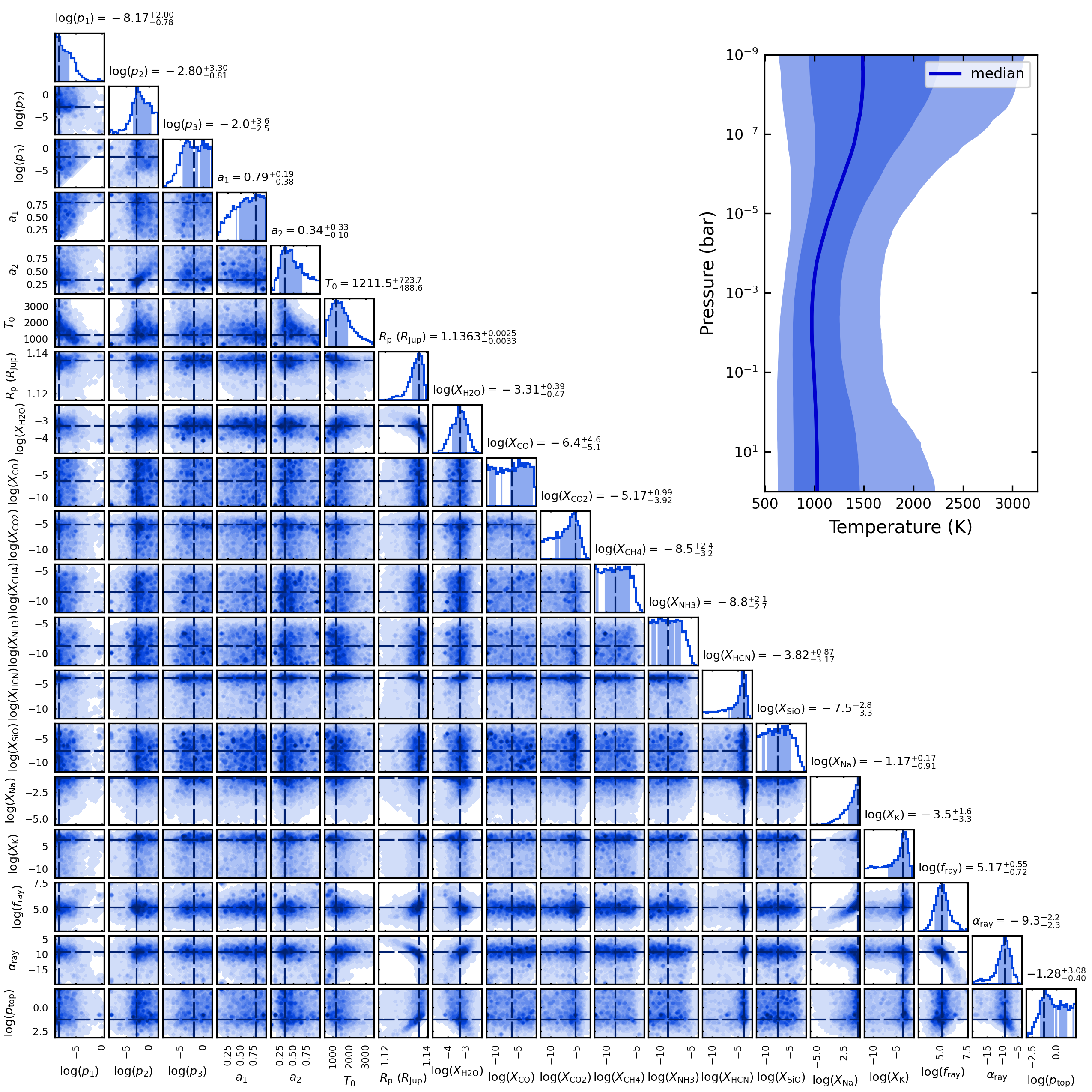}
\caption{Pairwise distribution (lower-left corner panels) and marginal
  histograms (diagonal) of the retrieved posterior parameters of the
  {\oneeightnineb} optical-IR retrieval.  The dashed lines indicate
  the marginal maximum likelihood of the parameters (the mode of the
  histograms, or median when the parameter not well constrained).  The
  shaded areas denote the 68\% highest posterior density.  The
  top-right panel shows the posterior distribution of the
  temperature-profile models (the median and the span of the 68\% and
  95\% central credible intervals).}
\label{fig:oir_posteriors}
\end{figure*}

Table \ref{table:oir_retrieval} shows the parameterization, priors, and
retrieved posterior values for the analysis of the {\oneeightnineb}
optical-IR transmission spectrum.  Figure \ref{fig:oir_posteriors}
shows the posterior distributions of the model parameters and
temperature profile.

{\renewcommand{\arraystretch}{1.4}
\begin{table}
\begin{minipage}{\linewidth}
\centering
\caption{HD 189733b Optical-IR Retrieval Parameters Summary}
\label{table:oir_retrieval}
\begin{tabular*}{\linewidth} {llr@{\hskip 0in}l}
\hline
Parameter     & Priors$^\dagger$ & \multicolumn2c{Retrieved value$^\ddagger$} \\
\hline
$\log_{10}(p_1/{\rm bar})$ & $\mathcal U(-9, 2)$   &  $-8.17$ & $^{+2.00}_{-0.78}$ \\
$\log_{10}(p_2/{\rm bar})$ & $\mathcal U(-9, 2)$   &  $-2.80$ & $^{+3.30}_{-0.81}$ \\      
$\log_{10}(p_3/{\rm bar})$ & $\mathcal U(-9, 2)$ and $p_3>p_1$ & $-2.0$&$^{+3.6}_{-2.5}$ \\
$a_1$ (K$^{-0.5}$)         & $\mathcal U(0.02, 2.0)$ & $0.79$ & $^{+0.19}_{-0.38}$  \\
$a_2$ (K$^{-0.5}$)         & $\mathcal U(0.02, 2.0)$ & $0.34$ & $^{+0.33}_{-0.10}$  \\ 
$T_0$ (K)                 & $\mathcal U(500, 3500)$ & $1210.0$ & $^{+720.0}_{-490.0}$  \\  
$R_{\rm planet}$ ($R_{\rm Jup}$) & $\mathcal U(0.5,2.0)$ & $1.1363$ & $^{+0.0025}_{-0.0033}$\\
$\log_{10}(X_{\rm H2O})$   & $\mathcal U(-12, -1)$ & $-3.31$ & $^{+0.39}_{-0.47}$  \\
$\log_{10}(X_{\rm CO})$    & $\mathcal U(-12, -1)$ & $-6.4$ & $^{+4.6}_{-5.1}$  \\
$\log_{10}(X_{\rm CO2})$   & $\mathcal U(-12, -1)$ & $-5.17$ & $^{+0.99}_{-3.92}$  \\
$\log_{10}(X_{\rm CH4})$   & $\mathcal U(-12, -1)$ & $-8.5$ & $^{+2.4}_{-3.2}$  \\
$\log_{10}(X_{\rm NH3})$   & $\mathcal U(-12, -1)$ & $-8.8$ & $^{+2.1}_{-2.7}$  \\
$\log_{10}(X_{\rm HCN})$   & $\mathcal U(-12, -1)$ & $-3.82$ & $^{+0.87}_{-3.17}$  \\
$\log_{10}(X_{\rm SiO})$   & $\mathcal U(-12, -1)$ & $-7.5$ & $^{+2.8}_{-3.3}$  \\
$\log_{10}(X_{\rm Na})$    & $\mathcal U(-12, -1)$ & $-1.17$ & $^{+0.17}_{-0.91}$  \\
$\log_{10}(X_{\rm K})$     & $\mathcal U(-12, -1)$ & $-3.5$ & $^{+1.6}_{-3.3}$  \\
$\log_{10}(f_{\rm ray}$)   & $\mathcal U(-4, 9)$   & $5.17$ & $^{+0.55}_{-0.72}$  \\
$\alpha_{\rm ray}$        & $\mathcal U(-20, 0)$  & $-9.3$ & $^{+2.2}_{-2.3}$  \\
$\log_{10}(P_{\rm top}/{\rm bar})$ & $\mathcal U(-7,2)$ & $-1.28$ & $^{+3.08}_{-0.40}$ \\
\hline
\end{tabular*}
\end{minipage}
\begin{tablenotes}
\item $^\dagger$ $\mathcal U(a, b)$ stands for a uniform distribution between $a$ and $b$.
\item $^\ddagger$ The values correspond to the maximum likelihood of
  the marginal posterior distributions. The uncertainties correspond
  to the span of the 68\% highest-posterior-density credible interval
  \citep{Andrae2010arxivErrorEstimation}.
\end{tablenotes}
\end{table}
}

\end{appendix}

\end{document}